\documentclass[fleqn,usenatbib,useAMS]{mnras}

\usepackage{graphicx}	
\usepackage{amsmath}	
\usepackage{amssymb}	
\usepackage{multicol}        
\usepackage{pdflscape}	
\usepackage{csvsimple}
\usepackage{rotating}
\usepackage{pdfpages}
\usepackage{url}
\usepackage{threeparttable}

\usepackage[T1]{fontenc}
\usepackage{ae,aecompl}

\usepackage{newtxtext,newtxmath}

\author[Kramer et al.]{\parbox{\textwidth}{ M.~Kramer$^{1,2}$\thanks{E-mail: mkramer@mpifr-bonn.mpg.de},
I.H.~Stairs$^{3}$, 
V. Venkatraman~Krishnan$^{1}$, 
P.~C.~C.~Freire$^{1}$, 
F.~Abbate$^1$, 
M.~Bailes$^{4,5}$, 
M.~Burgay$^6$, 
S.~Buchner$^7$, 
D.~J.~Champion$^1$, 
I.~Cognard$^{8,9}$,
T.~Gautam${^1}$, 
M.~Geyer$^7$, 
L.~Guillemot$^{9,8}$,
H.~Hu$^1$, 
G.~Janssen$^{10}$, 
M.~E.~Lower$^{4,11}$, 
A. Parthasarathy$^1$, 
A. Possenti$^{6,12}$,
S.~Ransom$^{13}$,
D.~J.~Reardon$^{4,5}$, 
A.~Ridolfi$^{6,1}$, 
M.~Serylak$^7$, 
R.~M.~Shannon$^{4,5}$, 
R.~Spiewak$^{2,4}$, 
G.~Theureau$^{9,8}$,
W.~van~Straten$^{14}$, 
N.~Wex$^1$, 
L.~S.~Oswald$^{15}$, 
B.~Posselt$^{15,16}$,
C.~Sobey$^{17}$,
E.~D. Barr$^1$, 
F.~Camilo$^7$, 
B.~Hugo$^{7,18}$, 
A.~Jameson$^{4,5}$, 
S.~Johnston$^{11}$, 
A.~Karastergiou$^{15}$, 
M.~Keith$^2$, 
S.~Os{\l}owski$^{4,19}$} \\ \\ \\ \\
\parbox{\textwidth}{$^1$ Max-Planck-Institut f\"{u}r Radioastronomie, Auf dem H\"{u}gel 69, D-53121 Bonn, Germany\\
$^2$ Jodrell Bank Centre for Astrophysics, University of Manchester, M13 9PL, UK\\
$^{3}$ Department of Physics and Astronomy, University of British Columbia, 6224 Agricultural Road, Vancouver, BC V6T 1Z1, Canada \\
$^4$  Centre for Astrophysics and Supercomputing, Swinburne University of Technology, P.O. Box 218, Hawthorn, VIC 3122, Australia\\
$^5$ ARC Centre of Excellence for Gravitational Wave Discovery (OzGrav)\\
$^6$ INAF - Osservatorio Astronomico di Cagliari, Via della Scienza 5, 09047 Selargius (CA), Italy\\
$^7$ South African Radio Astronomy Observatory, 2 Fir Street, Black River Park, Observatory 7925, South Africa\\
$^8$ Station de Radioastronomie de Nançay, Observatoire de Paris, CNRS/INSU, Université d'Orléans, 18330, Nançay, France \\
$^9$ Laboratoire de Physique et Chimie de l'Environnement, CNRS, 3A Avenue de la Recherche Scientifique, 45071, Orléans Cedex 2, France \\
$^{10}$ ASTRON, Netherlands Institute for Radio Astronomy, Oude Hoogeveensedijk 4, 7991 PD, Dwingeloo, The Netherlands\\
$^{11}$ CSIRO Astronomy \& Space Science, Australia Telescope National Facility, P.O. Box 76, Epping, NSW 1710, Australia \\
$^{12}$ Universit\'a di Cagliari, Dipartimento di Fisica, S.P. Monserrato-Sestu Km 0,700 - 09042 Monserrato (CA), Italy\\
$^{13}$ National Radio Astronomy Observatory, 520 Edgemont Rd., Charlottesville, VA 22903, USA\\
$^{14}$ Institute for Radio Astronomy \& Space Research, Auckland University of Technology, Private Bag 92006, Auckland 1142, NZ\\
$^{15}$ Oxford Astrophysics, University of Oxford, Denys Wilkinson Building, Keble Road, Oxford OX1 3RH, UK\\
$^{16}$ Department of Astronomy \& Astrophysics, Pennsylvania State University, 525 Davey Lab, 16802 University Park, PA, USA\\
$^{17}$CSIRO Astronomy and Space Science, PO Box 1130 Bentley, WA 6102, Australia \\ 
$^{18}$ Rhodes University: Department of Physics and Electronics, Rhodes University, Artillery Road, Grahamstown, South Africa\\
$^{19}$ Gravitational Wave Data Centre, Swinburne University of Technology, P.O. Box 218, Hawthorn, VIC 3122, Australia}\\
}

\title[The MeerKAT Relativistic Binary Programme]
{The Relativistic Binary Programme on MeerKAT: Science objectives and first results}

\date{Last updated; in original form}

\pubyear{2021}

\begin{document}
\label{firstpage}
\pagerange{\pageref{firstpage}--\pageref{lastpage}}
\maketitle

\begin{abstract}
We describe the ongoing Relativistic Binary programme (RelBin), a part of the MeerTime large survey project with the MeerKAT radio telescope. RelBin is primarily focused on observations of relativistic effects in binary pulsars to enable measurements of neutron star masses and tests of theories of gravity. We selected 25 pulsars as an initial high priority list of targets based on their characteristics and observational history with other telescopes. In this paper, we provide an outline of the programme, present polarisation calibrated pulse profiles for all selected pulsars as a reference catalogue along with updated dispersion measures. We report Faraday rotation measures for 24 pulsars, twelve of which have been measured for the first time. More than a third of our selected
pulsars show a flat position angle swing confirming earlier observations. We demonstrate the ability of the Rotating Vector Model (RVM), fitted here to seven binary pulsars, including the Double Pulsar (PSR J0737$-$3039A), to obtain information about the orbital inclination angle. We present  a high time resolution light curve of the eclipse of PSR~J0737$-$3039A by the companion's magnetosphere, a high-phase resolution position angle swing for PSR~J1141$-$6545, an improved detection of the Shapiro delay of PSR J1811$-$2405, and pulse scattering measurements for PSRs~J1227$-$6208, J1757$-$1854, and J1811$-$1736. Finally, we demonstrate that timing observations with MeerKAT improve on existing data sets by a factor of, typically, 2--3, sometimes by an order of magnitude. 
\end{abstract}



\clearpage

\section{Introduction}

Pulsars are remarkable laboratories for studying fundamental physics. When the rotation of a pulsar in a binary system is tracked with high precision using a technique called pulsar timing, we can study the orbit of the pulsar about the centre of mass that it shares with a companion object. If the orbit is compact enough, timing may reveal a number of relativistic effects that depend on the masses of the pulsar and its companion, apart from 
the Keplerian parameters of the orbit that are readily measured. Consequently, studying binary radio pulsars enables us to probe relativistic gravity as well as precisely measure masses of neutron stars (e.g. \citealt{Taylor&Weisberg1982,vanStraten2001,KramerEtAl2006,Weisberg&Huang2016}).
  
In both the highly relativistic interior and the vicinity of a pulsar (and its binary companion, in the cases of double neutron star systems or potential pulsar-black hole systems) space-time may significantly deviate from the predictions of General Relativity (GR; \citealt{DamourEspositoFarese1996}). Pulsar timing therefore is a rare tool for probing gravity in the (mildly-relativistic) strong-field regime, enabling high-precision tests of GR or alternative theories of gravity \citep{DamourTaylor1992,will18}. Perhaps best known precision tests of GR are the ones performed using the compact orbits of double neutron star (DNS) systems, such as the PSR B1913+16 \citep{Weisberg&Taylor1984} or the unique ``Double Pulsar'' \citep{KramerEtAl2006}. Binary pulsars with white dwarf companions enable tests of some of the fundamental properties of gravity such as a possible violation of the universality of free fall (e.g. PSR J0337$-$1715, the triple system with a pulsar and two white dwarfs, \citealt{ArchibaldEtAl2018,voisin2020}) or the existence of gravitational dipole radiation (e.g. PSR J1738+0333, \citealt{FreireEtAl2012}); both of these would result from the violation of the strong equivalence principle, a fundamental property of GR that is generally not incorporated in certain classes of alternative theories of gravity.

Investigating effects such as the Shapiro delay in tight binary orbits require time resolution, cadence and measurements with high sensitivity. Simultaneously, identifying the weak signatures of certain relativistic effects, such as gravitational wave damping, also needs long-term monitoring with good cadence. The new
MeerKAT telescope \citep{Camilo2018,MauchEtAl2020} with its excellent sensitivity is currently providing such observations as part of the MeerTime large science programme (LSP), which aims to precisely time radio pulsars and probe its various applications \citep{BailesEtAl2020}.

The Relativistic Binary (``RelBin'') theme of MeerTime LSP intends to make use of MeerKAT's sensitivity 
to not only improve on the existing tests of gravity but also measure new effects and probe new phenomena. As an example, we refer to observations of the Double Pulsar (\citealt{KramerEtAl2006,huhu2020}), where we can expect the precision of tests of gravity to go beyond the current best weak-field tests in the Solar System. In the case of the either Double Pulsar or PSR~J1757$-$1854, we expect that our measurements will provide one of the first measurements of the  moment of inertia of a neutron star \citep[see ][and references therein]{KehlEtAl2016,CameronEtAl2018,huhu2020}. This information will provide an important handle on the equation of state (EOS) of neutron matter at densities above those of the atomic nucleus, a fundamental problem in astrophysics and nuclear physics \citep{huhu2020}. Precise measurements of neutron star (NS) masses by themselves provide important constraints on the EOS.
In particular, the discovery of massive neutron stars  (eg. \citealt{DemorestEtAl2010, AntoniadisEtAl2013,CromartieEtAl2020}) rules out a variety of soft EOS (see e.g.~\citealt{Ozel&Freire2016}), and it
suggests that there is a large population of massive neutron stars; these large masses are likely a birth property \citep{tauris17}. As our knowledge about the NS mass distribution improves with RelBin observations, we will get closer to identifying the maximum possible mass for a neutron star that needs to be sustained by any proposed EOS. We will also be in a position to determine whether the NS mass distribution is bi-modal, as suggested by \cite{AntoniadisEtAl2016}. 

Determining the mass of the companion at the same time also allows us also to test theories of binary evolution (e.g. \citealt{TaurisAndSavonije1999}). RelBin observations will also result measurement of improved astrometric information (parallaxes, proper motions and hence velocities) for millisecond and binary pulsars, where our observations are complemented by general millisecond pulsar timing in MeerTime (Spiewak et al. in prep.). Such information will allow us to infer birth velocities and to constrain asymmetric supernova mechanisms, particularly in DNS systems \citep[see e.g.][]{std+06,Tauris+2017}.

The aim of this paper is to describe the RelBin science programme and to complement the recent description of other MeerTime themes  \citep[e.g.][]{jkk+20}. We lay out the associated science questions and provide not only first results of the programme, but we also aim to provide guidance for the community which results are to be expected and how to steer similar observing ideas towards realisation while complementing the RelBin presented and expected. We do this  by providing an overview of the list of sources that are currently studied, present early results and predictions for the future of MeerTime's RelBin project. In addition, we conduct a ``mini-census'' of the selected sources to help quantifying future outcomes.

\section{The Relativistic Binary Programme} 
\label{sec:RelBinProgramme}

Pulsar timing registers the pulse arrival times of pulsar signals at Earth and transforms these topocentric times of arrival (ToAs) via a hypothetical ToA at the Solar System Barycentre (barycentric ToAs) into the reference frame of the pulsar. This allows us to precisely count the number of rotations of the neutron star with the help of a timing model. Using this, we can measure the relevant timing parameters, the precision of which increases with the number and precision of the ToAs as well as the total time span of the observations \citep{Lorimer&Kramer2005}. If the pulsar has a companion, the impact on the ToAs from the corresponding orbital motion can be described, in the simplest case, with five Keplerian parameters. Deviations from a simple Keplerian orbit due to astrophysical or relativistic effects can be described by the addition of theory-independent phenomenological ``post-Keplerian'' (PK) parameters \citep{DamourTaylor1992}. For any metric theories of gravity, these PK parameters can be written as functions of the well-measured Keplerian parameters and the unknown masses of the pulsar and its companion. Measuring $n_{\rm PK}=2$ PK parameters thereby provides a means to measure these masses for a given theory of gravity. If more than two PK parameters can be measured, $n_{\rm PK}>2$, the set of available equations is over-determined. In this case, one can check for the self-consistency of a theory of gravity, allowing $n_{\rm PK}-2$ independent tests of the theory. 

The PK parameters, once determined, often also provide access to additional information about the system. For instance, a comparison of the observed orbital decay with the value expected from GR from gravitational wave damping may reveal kinematic effects that allow precise distance measurements (e.g.~\citealt{Bell&Bailes1996,stairs2002,smits2011}). Similarly, PK parameters may provide information about the system geometry (e.g.~\citealt{Kramer1998,vanStraten2001}). In some cases, they reveal information about the spin
of the components, as in the case of PSR~J1141$-$6545, where the fast spin of the companion white dwarf to the pulsar was detected from the change of the orbital inclination of the system \citep{VenkatramanKrishnanEtAl2020b}.
As explained in detail in \cite{huhu2020}, such instances of spin-orbit coupling may provide information on the mentioned moment-of-inertia of neutron stars. 

We summarise the sources identified as prime targets for our coordinated programme in
Table~\ref{tab:expectations}. We divide the sources into two classes. The first class has already masses determined for the binary components, so that we can estimate which PK parameters should
be newly measurable or can be improved for tests of theories of gravity. The second class are binaries  
that are likely candidates for mass measurements via relativistic effects, based on their known orbital configuration and the achievable timing precision. Currently, however, for most of these latter sources only the mass function is available so far. Since the outcome of our
observations will depend on the actual orbital inclination angle that is yet unknown, 
simulations of the possible precision to be achieved are of limited use. Furthermore, for many
sources less than a year of timing data is available, so that post-fit timing residuals 
may not be fully representative yet. Instead, we opt to 
express our likely ability to succeed by  a comparison between the previous currently best available ToA precision and the one based on our early RelBin results presented here. 
To ease the comparison, we
base it on a uniform set of
observing parameters such as same receiver package, bandwidth, integration time and
standard processing pipeline.
The values shown in Table~\ref{tab:expectations} are compared to published values adjusted for the same integration time.

In order to achieve our goals, the strategy is to conduct observations with monthly cadence.
Along with this, depending on the orbital period, we either conduct
full orbit observations ($P_{\rm b}\sim$ few hours) and  orbital campaigns with cadence
of a few hours for longer orbits. For orbits of a few days (e.g.~PSR J1017$-$7156),
we perform long (i.e.~few hour)  observations over superior conjunction followed by 
shorter observations 
on consecutive days to cover eventually the whole orbit. For very long orbital period pulsars 
(e.g.~PSR J0955$-$6150) we schedule a dedicated observation close to superior conjunction, while
the coverage for the rest of the orbit comes from regular (i.e. monthly) cadence
observations.

\subsection{Tests of relativistic gravity}
\label{sec:gravity}

In order to obtain $n_{\rm PK}>2$ PK parameters required to test theories of gravity, the pulsar orbit usually needs to be sufficiently compact. In RelBin we concentrate on sources that can benefit from the excellent sensitivity of MeerKAT and are easily accessible (i.e. in the southern hemisphere), namely PSR J0737$-$3039A/B (i.e. the Double Pulsar) and PSRs J1141$-$6545, J1756$-$2251,
J1757$-$1854 and J2222$-$0137. All but PSR J1141$-$6545 and PSR J2222$-$0137 are DNS systems. These and the other sources will be discussed in detail in Section~\ref{sec:specific_sources}, where we present the first MeerKAT observations and results. In subsequent work, we will combine our MeerKAT data with those obtained at other telescopes to expand the timing baseline (see also Section~\ref{sec:parkes}). In all cases, we can expect that the higher precision obtained with MeerKAT will lead to significant improvements to the overall timing measurements, especially the PK parameters (see Section~\ref{sec:predictions}).

\subsection{Mass measurements of neutron stars}

Our objectives are both to improve on existing mass determinations and to derive new measurements. The binary systems considered for mass measurements can then be split into the following categories:
\begin{itemize}
    \item[-] Pulsars in nearly circular orbits with likely white dwarf (WD) companions, for which we expect to obtain Shapiro delay measurements.
    \item[-]  Millisecond pulsars (MSPs) that are also timed as part of MeerTime in a ``Pulsar Timing Array'' (PTA) programme (Spiewak et al. in prep.). The RelBin observations complement the regular PTA monitoring by dedicated sessions aimed on optimising orbital phase coverage. Shapiro delay measurements obtained in this way may be combined with potential measurements of the PK parameters $\dot{\omega}$ and $\dot{x}$, i.e.~periastron advance and change in the projected semi-major axis, respectively. These effects cannot only be caused by relativistic gravity but, for instance, by kinematic effects.
    \item[-] Eccentric MSPs, where we will be able to measure periastron advance, $\dot\omega$, and Shapiro delay (J0955$-$6150 \& J1618$-$3921).
    \item[-] Other pulsars where the companion has yet to be identified as either a WD or NS. For these systems (e.g.~PSRs J1802$-$2124, J1811$-$1736, J1930$-$1852) we expect to measure periastron advance, Shapiro delay and eventually gravitational wave damping, i.e.~the PK parameter $\dot{P}_{\rm b}$.
\end{itemize}

\begin{table*}
    \caption{
    Basic parameters of the sources in the MeerKAT Relativistic binary programme. We list the spin and orbital periods, the eccentricity and the mass function. We also list the initial Time-of-Arrival (ToA) measurement precision achieved for all the pulsars with a 2048 second, full-band  integration. The values are 
    obtained using a standard processing pipeline with a single template for the whole band. Hence these estimates are conservative, unless pulsar was highly scintillating up. We compare this precision with the currently best available TOA precision (using values to the best of our knowledge based on literature and our own data sets regardless of observing frequencies, see last column), adjusted for the same integration time as here. For Southern-sky sources, these improvements are about one order of magnitude. The penultimate column lists the science goals that we expect to achieve with our observations. See Table footnotes and text for more details.
    }

\begin{center}
\begin{threeparttable}
\begin{tabular}{|l|r|r|r|c|c|c|r|}
\hline
\hline
PSR Name & Spin period, & Orbital period, & Eccentricity,& Mass Function, & TOA Precision  & Science Goals & Ref. \\
& $P_0 ({\rm ms})$ & $P_{\rm b}$ (days) & $e$ & $M_{\rm f}(M_{\odot})$  & ($\mu$s)/ improvement &  & \\
\hline
J0737$-$3039A & 22.699 & 0.102 & 0.08778  & 0.291  & 0.65 / $>2\times$ & S, GW, EOS, E, LB,DTH, SO & (1) \\
J0955$-$6150 & 1.999 & 24.578 & 0.11 & 0.0042 &  0.43 / $>20\times$ & OM, S, M & (2) \\
J1017$-$7156 & 2.339  & 6.511  & 0.00014 & 0.0029  & 0.05 / $>3\times$& OM, S, M & (3)$\dagger$\\
J1141$-$6545 & 393.899  & 0.199  & 0.17189  & 0.1766  & 4.55 / $>2\times$ &S, SO, M & (4)$\dagger$\\
J1157$-$5112 & 43.589  & 3.507 & 0.00040 & 0.2545 & 3.27 / $>10\times$ & OM, S, M & (5)\\
J1227$-$6208 & 34.528 & 6.721 & 0.00115  & 0.2968  & 1.67 / $>5\times$& OM, S, M & (6)\\
J1435$-$6100 & 9.348  & 1.355  & 0.00001  & 0.1383  & 0.35 / $>10\times$ & S, M & (7) \\
J1454$-$5846 & 45.249 & 12.423  & 0.00190  & 0.1299  & 6.48 / $>8\times$ & OM, S, M & (7) \\
J1528$-$3146 & 60.8222 & 3.180 & 0.00021 & 0.1595  &0.37 / $>7\times$ & OM, S, M & (8)$\dagger$ \\
J1603$-$7202 & 14.842  & 6.309  & 0.00001  & 0.0088  & 0.27 / $>3\times$ &S, GW, INC  & (3)$\dagger$ \\ 
J1618$-$3921 & 11.987 & 22.746  & 0.02741 & 0.0023  & 1.51 / $>20 \times$ & OM, S, M & (9) \\
J1727$-$2946 & 27.083 & 40.308  & 0.04563  & 0.1194  &2.21 / $> 10 \times$  & OM, S, M  & (10) \\
J1732$-$5049 & 5.313 & 5.263  & 0.00001  & 0.0025  &0.40 / $>3\times$ & S, M  & (3)$\dagger$ \\
J1748$-$2021B & 16.760 & 20.550  & 0.57016 & 0.0002  &10.73 / $\sim2\times$& OM, S, M & (11)$\dagger$ \\
J1753$-$2240 & 95.138 & 13.638 & 0.30358  & 0.0343  &48.33 / $> 10 \times$ & OM, S, M & (12) \\
J1756$-$2251 & 28.462 & 0.320  & 0.18057  & 0.2201  &1.69 / $> 1.7 \times$ & OM, S, GW, M & (13) \\
J1757$-$1854 & 21.497 & 0.184  & 0.60581 & 0.3572  &10.41 / $>1.2 \times$ & OM, GW, S, M, DTH, EOS  & (14)\\

J1757$-$5322 & 8.870  & 0.453  & $<0.00001$  & 0.0475  &0.39 / $>10\times$ & GW, S, M & (5)\\
J1802$-$2124 & 12.648 & 0.699  & $<0.00001$ & 0.1131  &0.26 / $> 2.5 \times$ & GW, S, M & (15)\\
J1811$-$1736 & 104.182  & 18.779  & 0.82801 & 0.1281  &45.03 / $> 7 \times$ & OM,S, M & (16) \\
J1811$-$2405 & 2.661  & 6.272  & $<0.00001$ & 0.0051  &0.18 / $> 2.7 \times$ &S, M  & (17)\\
J1930$-$1852 & 185.520 & 45.060  & 0.39886 & 0.3469  &12.43 / $> 1.5 \times$ & OM, S, M & (18)\\ 
J1933$-$6211 & 3.543  & 12.819  & $<0.00001$  & 0.0121 & 0.32 / $>3\times$ &S, M & (19)$\dagger$  \\
J2129$-$5721 & 3.726  & 6.625  &  0.00001  & 0.0010 & 0.15 / $>3\times$ &S, M & (3)$\dagger$ \\
J2222$-$0137 & 32.818 & 2.446  &  0.00038  & 0.2291  &1.29 / $> 1.05 \times$ &OM,S,GW, M & (20)\\

\hline
\hline
\label{tab:expectations}
\end{tabular}
\begin{tablenotes}
\item
\small Abbreviations of science goals:  	 
	 	 S - Shapiro delay measurement,
	 	 OM - Periastron advance via $\dot\omega$,
	 	 GW - Gravitaional wave damping via $\dot{P}_{\rm b}$,
	 	 M - Mass measurements,
	 	 SO - spin-orbit coupling (classical and relativistic spin and orbital precession),
	 	 EOS - Equation of state,
	 	 INC - constraints on inclination from proper motion,
	 	 LB - Light - Bending effect,
	 	 DTH - relativistic deformation of the orbit via $\delta_\theta$,
	 	 E - Eclipse studies.\\
	 	 The references for the previously achieved ToA precision are (1) \cite{huhu2020}, (2) \cite{CamiloEtAl2015}, (3) \cite{KerrEtAl2020}, (4) \cite{VenkatramanKrishnanEtAl2020b}, (5) \cite{EdwardsAndBailes2001}, (6) \cite{BatesEtAl2015}, (7) \cite{CamiloEtAl2001}, (8) \cite{jbo+07},  (9) \cite{ocg+18}, (10) \cite{lem+15},  (11) \cite{FreireEtAl2008},
	 	 (12) \cite{KeithEtAl2009}, (13) \cite{FerdmanEtAl2014}, (14) \cite{CameronEtAl2018}
	 	 (15) \cite{fsk+10}, (16) \cite{CorongiuEtAl2007} (17) \cite{NgEtAl2020}, (18) \cite{srm+15}, 
	 	 (19) \cite{GraikouEtAl2017}, (20) \cite{cognard2017}. For pulsars marked with $\dagger$ we compare the MeerKAT ToA precision to the best data from other telescopes that we have access to; these are significantly better than the data in the latest published reference.
\end{tablenotes}
\end{threeparttable}
\end{center}
\end{table*}

\begin{table*}
\caption{Measurements of DM and RM$_{\rm meas}$ for the RelBin list of sources. The measurements are obtained from the longest observation of the pulsar performed in the course of Relbin. In order to derive at the rotation measure of the pulsar, RM$_{\rm PSR}$, we correct for ionospheric contributions, RM$_{\rm Iono}$, which are computed using \textsc{ionFR} \citep{Sotomayor-BeltranEtAl2013}.  We provide the RM values from \textsc{psrcat} \citep{ManchesterEtAl2005} where available. }
	 \begin{threeparttable}
	 	 \begin{tabular}{|l|c|c|c|c|c|c|}
	 	 \hline
	 	 \hline
	 	 PSR Name  & Observation  & DM & $\rm RM_{meas}$  & $ \rm RM_{\rm iono}$ & $ \rm RM_{\rm PSR}$  & $\rm RM_{\rm psrcat}$\\
	 	 & epoch, (MJD) & ($ \rm pc~cm^{-3}$) & ($ \rm rad~m^{-2}$) & ($ \rm rad~m^{-2}$) & ($ \rm rad~m^{-2}$) & ($ \rm rad~m^{-2}$)\\
	 	 \hline
	 	 J0737$-$3039A & 58783.3 & 48.92(1) & 120.22(2) & $-$0.6(2) & 120.84(20) & $112(2)$ \\
	 	 J0955$-$6150 & 58836.9 & 160.906(1) & $-$48(5) & $-$0.6(2) & $-$47(5) & $--$\\
	 	 J1017$-$7156 & 58775.5 & 94.216(1) & $-$63.8(1) & $-$1.54(5) & $-$62.29(12) & $-63(1)$ \\
	 	 J1141$-$6545 & 58576.0 & 116.2(2) & $-$92.6(2) & $-$0.33(5) & $-$92.25(25) & $-93(3)$ \\
	 	 J1157$-$5112 & 58698.5 & 39.78(8) & $-$5.3(3) & $-$0.40(8) & $-$4.93(31) & $-33(14)$\\
	 	 J1227$-$6208 & 58806.0 & 362.96(2) & 48.0(8) & $-$0.57(6) & 48.6(8) & $--$ \\
	 	 J1435$-$6100 & 58860.9 & 113.788(5) & $-$53.1(2) & $-$0.6(2) & $-$52.49(22) & $--$ \\
	 	 J1454$-$5846 & 58849.4 & 116.10(2) & 82(5) & $-$1.3(4) & 84(5) & $--$\\
	 	 J1528$-$3146 & 58829.4 & 18.17(3) & $-$20.00(7) & $-$1.0(2) & $-$19.00(21) & $-32(14)$\\
	 	 J1603$-$7202 & 58986.8 & 38.055(6) & 30.15(9) & $-$0.38(4) & 30.53(10) & $35(2)$\\
	 	 J1618$-$3921 & 58861.2 & 117.942(6) & 176.8(1) & $-$0.8(3) & 177.58(35) & $--$\\
	 	 J1727$-$2946 & 58984.0 & 60.72(1) & $-$93(1) & $-$0.23(7) & $-$93.2(1.5) & $-61(32)$\\
	 	 J1732$-$5049 & 58986.8 & 56.823(2) & $-$8.8(2) & $-$0.33(6) & $-$8.43(24) & $-9(7)$\\
	 	 J1748$-$2021B & 58694.7 & 220.97(1) & $-$1.7(7) & $-$0.13(6) & $-$1.6(7) & $--$\\
	 	 J1753$-$2240 & 58984.0 & 158.42(9) & 0.0(0) & $-$0.19(5) & 0.19(5) & $--$\\
	 	 J1756$-$2251 & 58924.2 & 121.27(1) & $-$10.1(1) & $-$0.4(3) & $-$9.69(32) & $10(11)$\\
	 	 J1757$-$1854 & 58591.1 & 378.25(1) & $-$700(1) & $-$0.15(8) & $-$700.3(1) & $--$\\
	 	 J1757$-$5322 & 58868.4 & 30.799(4) & 69.4(2) & $-$0.8(1) & 70.21(20) & $--$\\
	 	 J1802$-$2124 & 58986.9 & 149.601(6) & 294.5(2) & $-$0.22(5) & 294.75(17) & $286(14)$\\
	 	 J1811$-$1736 & 58810.4 & 474.4(2) & $-$190(17) & $-$0.78(3) & $-$188(17) & $--$\\
	 	 J1811$-$2405 & 58750.6 & 60.615(1) & 30.3(2) & $-$0.5(3) & 30.79(29) & $23(3)\dagger$\\
	 	 J1930$-$1852 & 58986.9 & 42.93(8) & 9(2) & $-$0.25(5) & 9.3(2) & $--$\\
	 	 J1933$-$6211 & 58746.8 & 11.521(2) & 9.2(1) & $-$0.32(4) & 9.57(15) & $--$\\
	 	 J2129$-$5721 & 58752.0 & 31.846(2) & 22.50(6) & $-$0.38(6) & 22.88(9) & $22.3(3)$\\
	 	 J2222$-$0137 & 58752.0 & 3.28(2) & 2(1) & $-$0.14(3) & 2.0(1) & $2.6(1.0)$\\
	 	 \hline
	 	 \end{tabular} \label{table:DMRM}
	 \begin{tablenotes}
	 	 \item \small $\dagger$ \textsc{psrcat}'s values for this pulsar were not up to date. The latest estimates come from \citep{NgEtAl2020} and is 21(9)\, $ \rm rad~m^{-2}$ which is consistent with our measurements.
	    \end{tablenotes}
\end{threeparttable}
\end{table*}

\section{Observations and data analysis}

\label{sec:obs_and_data_analysis}

We provide here a brief introduction to the telescope and the backend system used for the MeerTime project. For further details on the instrumentation for pulsar observations, see  \cite{BailesEtAl2020}.

\subsection{The MeerKAT telescope, calibration \& data acquisition system}
\label{sec:DAS}
The MeerKAT telescope is a 64-dish interferometer situated in  Karoo  region of South Africa and is operated by the South African Radio Astronomy Observatory (SARAO). Each dish is 13.9 m in diameter and currently has two operational receivers in its focus, in an offset-Gregorian configuration, with a gain of 2.8 K/Jy. The first ``L-band'' receiver operates with a bandwidth of 856 MHz centred at a frequency of 1284 MHz, and it is the instrument used for most of the pulsar observations presented in this paper.  The receiver has a very low system temperature of $\sim18$~K, making it one of the most powerful interferometers around 1.4 GHz. The second ``UHF-receiver'' is centred at a frequency of 816 MHz with a bandwidth of 544 MHz. It has recently been installed with commissioning and testing observations underway. A third receiver suite, the ``S-band'' receivers (operating at 1.75--3.5 GHz) have been designed and built by the Max-Planck-Institut für Radioastronomie (MPIfR). The receivers are currently being delivered to the telescope and commissioned upon arrival. Given that S-band observations promise significant improvement in timing precision for a number of sources (as detailed further below), RelBin will make extensive use of the MPIfR S-band system once it is fully operational in the next 12--18 months.

The signals from all 64 antennas are first amplified by chain of radio frequency amplifiers, and are sampled at radio frequency (no down-conversion) to produce complex voltage streams that are sent to the correlator-beamformer engine (CBF) via a 40-Gbps switch. The CBF coherently adds the voltages, channelizes to either 1024 or 4096 channels, beamforms up to 4 tied-array beams (TAB) and streams the channelized time series to the Pulsar Timing User Supplied Equipment (PTUSE) machines. The steps in the observing procedure as follows.

Before every pulsar observing session, calibration observations are performed to phase up the array and obtain polarisation calibration solutions. This includes observations of a calibrator source with well characterised flux and polarisation for a wide range of frequencies. During these observations, signal from a noise diode is injected into the voltage stream just before RF amplification in every antenna. The sources often used are PKS J0408$-$6545, PKS J0825$-$5010 and PKS J1939$-$6342 \citep{reynolds94,hugo2018a,hugo2018b}. The calibration procedure consists of multiple stages where instrumental delays as well as geometric delays are applied for each antenna and polarisation. The noise diode signal is used to calculate cross-polarisation delays and phases per antenna per polarisation. The next stage is to remove the bandpass response of the receivers. All the aforementioned corrections are incorporated into the complex gain (frequency, antenna and polarisation) which is applied post-channelisation, but prior to beamforming. This is done by the real-time pipeline which derives and stores the solutions in the telescope metadata system.  Note that for all the observations done before April 2020, polarisation calibration was done offline as the aforementioned cross-polarisation phase was not calculated by the calibration pipeline.  More details on polarisation calibration are provided by \cite{sjk+20}.

There are 4 PTUSE machines, each processing one beam at a time. On each PTUSE machine, the data streamed from the CBF are added to a \textsc{psrdada}\footnote{\url{ https://psrdada.sourceforge.net}} ring buffer in the CPU memory. The data from this ring buffer is asynchronously processed by the pipelines in the \textsc{dspsr}\footnote{\url{http://ascl.net/1010.006}} software library \citep{vanStraten2011}. The pipeline either records full-Stokes, search mode data at a sampling time of  9.57$\mu s$ (which can be configured to scrunched down up to 38$\mu s$), or folds the data (i.e. computes the phase-resolved average of the polarised flux) at the topocentric pulse period with 1024 phase bins and 8-s integration lengths. Both search and fold-mode acquisition can be configured to run with or without coherent dedispersion. Data shown here are obtained in coherent dedispersion mode.

\subsection{Data Analysis}
\label{sec:analysis}

The pulsar fold-mode and search-mode data from the PTUSE machines are periodically transferred to the OzStar supercomputing cluster at Swinburne University of Technology in Australia. The fold-mode archives are fed through a pipeline (\textsc{meerpipe}) which performs automated RFI excision and polarisation calibration. The RFI excision is performed using a modified version of \textsc{coastguard} \citep{LazarusEtAl2016} and the Jones matrices for polarisation calibration are obtained from the phase up observation of the telescope (see \ref{sec:DAS}). The cleaned, calibrated files are decimated to the required time and frequency resolution, and the Times of Arrival (ToAs) of the pulses are obtained using the \textsc{pat} programme in the \textsc{psrchive}\footnote{\url{http://ascl.net/1105.014}} software suite \citep{HotanEtAl2004}.

We used the longest observation taken on each pulsar to compute the updated dispersion and rotation measures (DMs and RMs) provided in Table \ref{table:DMRM}. The DM of the pulsar was obtained using the \textsc{pdmp} program that provides the dispersion measure that maximises the signal to noise ratio. The RM was obtained using the \textsc{rmfit} program. \textsc{rmfit} obtains the best RM by brute-force searching for the maximum signal-to-noise ratio for the linearly polarised flux, $L = \sqrt{Q^{2} + U^{2}}$, as a function of trial RM. The range of RMs trialled and the step size is  automatically determined such that the change in position angle (P.A.) over the band, P.A.$ < 1$ radian.  Both \textsc{pdmp} and \textsc{rmfit} are part of the  \textsc{psrchive} software package.

\subsection{Supporting Parkes observations}
\label{sec:parkes}

Observing cadence, significant orbital coverage and long timing baseline are crucial for the measurement of several relativistic effects and the masses of the component stars. In order to achieve this, we note that the RelBin programme on MeerKAT reported here is supported by further dedicated observing campaigns. In particular, we have 
an ongoing support project with the Parkes radio telescope where a subset of our pulsars is timed with the Ultra Wide-Band Low (UWL) receiver (Project ID P1032; PI Venkatraman Krishnan). These observations will not only help with gaining better orbital coverage but the ultra wide bandwidth of 4~GHz will also help in obtaining better constraints on orbital and temporal dispersion measure variations of the pulsar, an effect known to bias estimates of the relativistic parameters. These data will be included in subsequent publications on timing results for specific sources.

\section{The RelBin Sources}

We present the first results of our RelBin observations with MeerKAT by first reviewing the sample properties, before commenting on the specific sources in the next section.

\subsection{Pulse profiles}

\begin{figure*}
    \centering
    \includegraphics[width=0.9\textwidth]{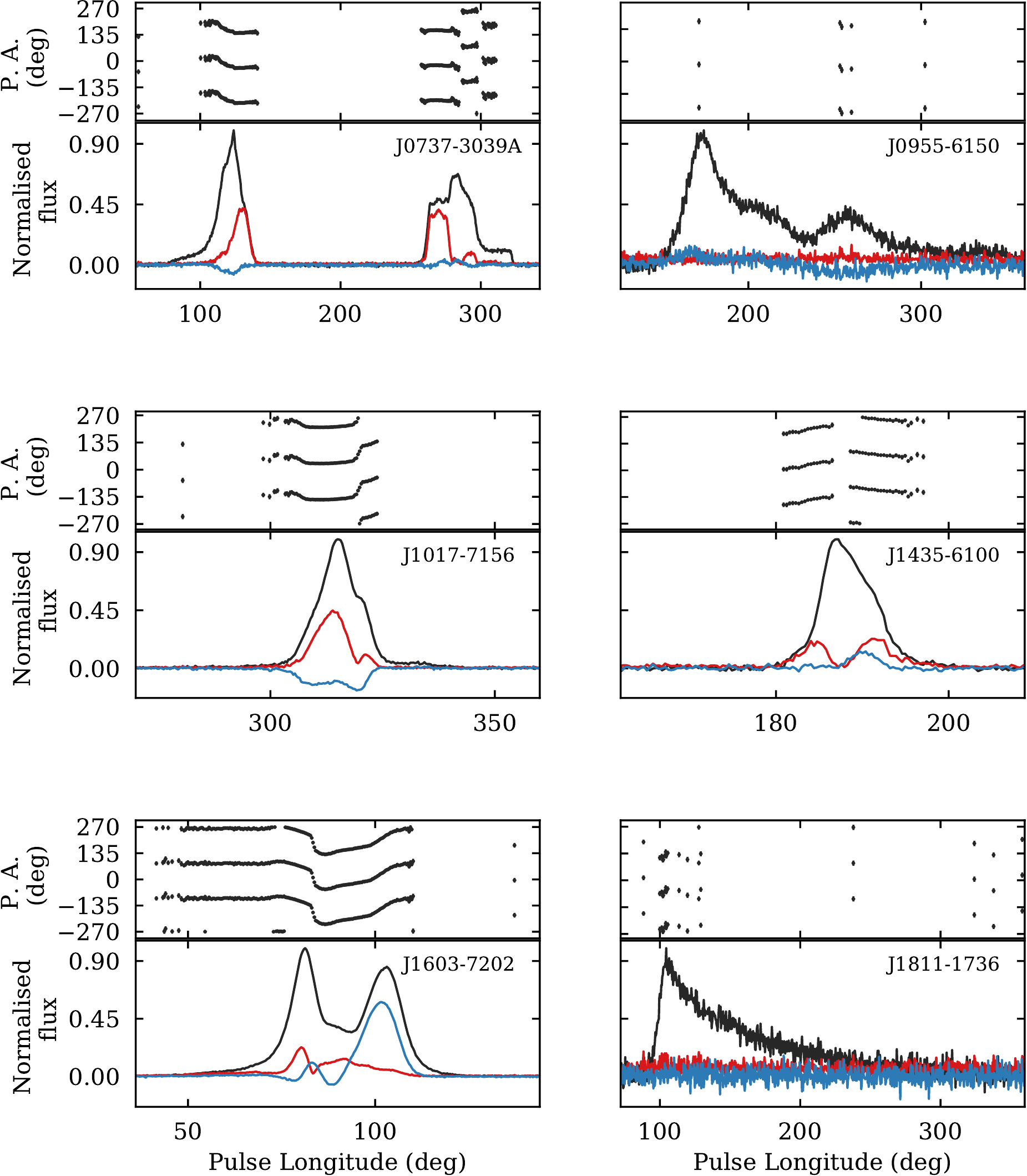}
    \caption{ {\color{black} Pulse profiles for the relativistic binaries with a complex P.A. swing or where the
    P.A. swing is not well defined.  Apart from PSR J0737$-$3039A, for reasons explained in the text, we do not attempt to fit a Rotating Vector Model (RVM) to these sources.
    All profiles were obtained with the L-band system, which is sensitive between 856 and 1712 MHz. Each pulsar was observed for a duration of 2048 seconds and folded at its topocentric period with 1024 bins across its rotational phase after coherently dedispersing at the best DM and summing all frequency channels. In each sub-plot, the bottom panel shows the normalised flux of the total intensity, linear and circular polarisation profiles plotted as black, red and blue respectively. The top panel shows the corresponding position angle of the linear polarisation. Several cycles of the P.A. are shown for clarity. The plots are zoomed in to only show the on-pulse regions. All position angles have been corrected for Faraday rotation and hence rotated to infinite frequency.}}
    \label{fig:complexPAa}
\end{figure*}

\begin{figure}
    \centering
    \includegraphics[width=0.4\textwidth]{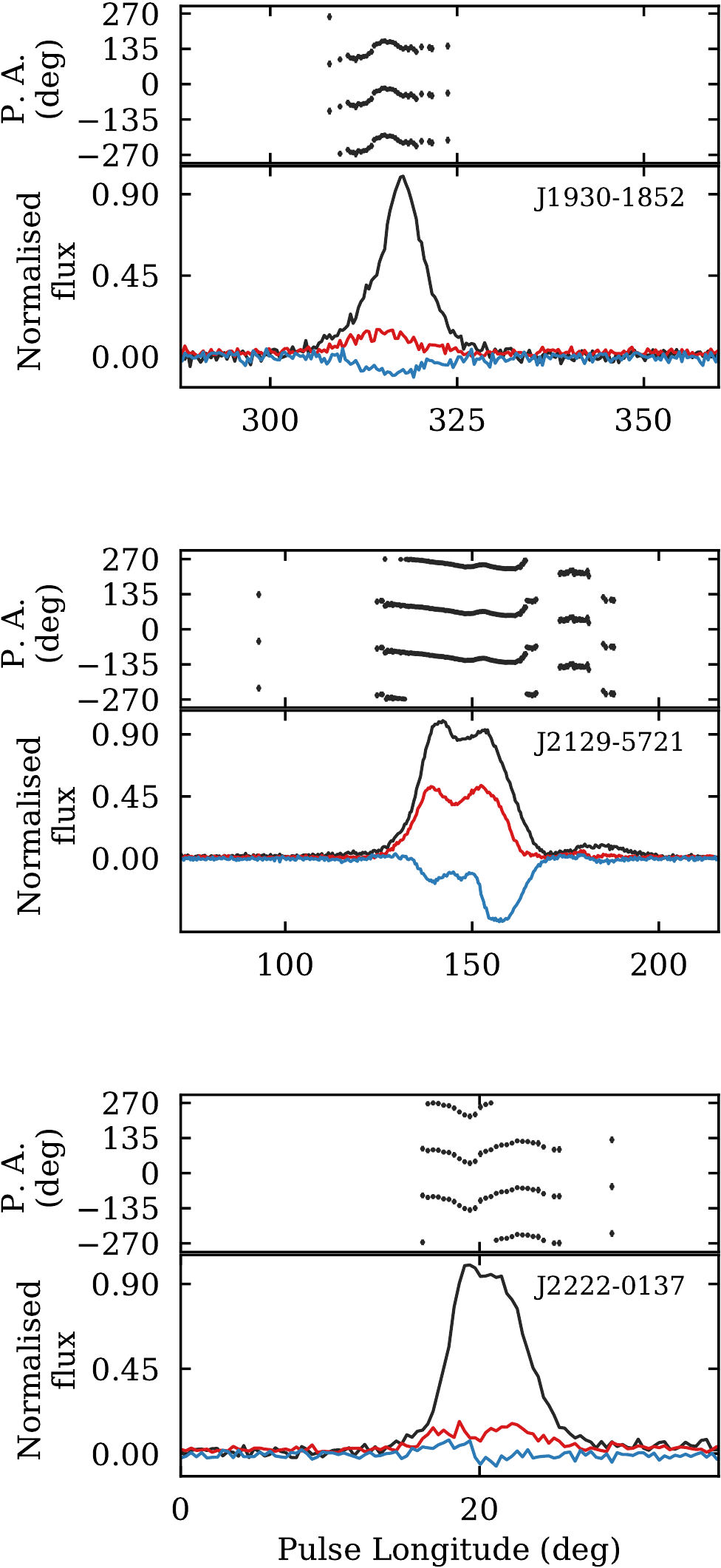}
    \caption{As Fig. \ref{fig:complexPAa} continued.}
    \label{fig:complexPAb}
\end{figure}

Figures \ref{fig:complexPAa}, \ref{fig:complexPAb}, \ref{fig:flatPAa},  \ref{fig:flatPAb},
\ref{fig:rvmPA}, \ref{fig:0737PA}
and \ref{fig:J1141-6545} 
show the L-band polarisation profiles and the corresponding position angle swings of their linear polarisation for all the pulsars. In order to demonstrate the relative brightness of the sources, and hence the expected timing precision, all profiles shown in Figs.~\ref{fig:complexPAa} - \ref{fig:rvmPA}
have been obtained with the same observing length of 2048 seconds. All profiles are well resolved, but those of
PSRs J1811$-$1736 and J1227$-$6208 are clearly affected by interstellar scattering. The same is true for PSR J1757$-$1854 at the lower part of our frequency band. In Section~\ref{sec:outlook} we will discuss the prospects of observing these three particular pulsars at higher frequencies, i.e.~using the S-band receivers, in more detail. Concentrating on the other pulsars, we notice a variety of pulse shapes, from relatively simple profiles as usually observed in non-recycled pulsars (cf.~\citealt{Lorimer&Kramer2005},
see e.g.~PSRs J1756$-$2251, J1930$-$1852) to complex profiles (e.g.~PSR J1933$-$6211) to rather unusual looking profiles (e.g.~PSRs J1454$-$5846 \& J1757$-$5322). Earlier studies have shown that recycled pulsars often have additional low-level pulse or interpulse components \citep{KramerEtAl1998}; these 
can also be seen in this sample (see e.g.~PSRs J1528$-$3146 \& J1811$-$2405). Whether this is caused by geometrical effects or extra-emission components, for instance from other outer regions in the otherwise rather compact magnetosphere as speculated before \citep{KramerEtAl1998,XilourisEtAl1998}, can be potentially addressed by inspecting the polarisation properties and in particular the shape of the position angle swing, as we do in the following.

\subsection{Updates on DM and RM}

Our measurements of DM and RM are obtained from the longest observation of each pulsar taken over the last year. The updated values along with the epoch of the observation are provided in Table  \ref{table:DMRM}, including new RM measurements for 12 pulsars. We use the publicly available software package, \textsc{ionFR}\footnote{\url{http://ascl.net/1303.022}}\citep{Sotomayor-BeltranEtAl2013}, to obtain the ionospheric Faraday rotation measure ($\rm RM_{Iono}$) to the measured RMs ($\rm RM_{meas}$) corresponding to each observation  \citep{Sotomayor-BeltranEtAl2013}. The software uses inputs from the International Geomagnetic Reference Field\footnote{\url{https://www.ngdc.noaa.gov/IAGA/vmod/igrf.html}} and the International GNSS service vertical total electron content maps\footnote{\url{ftp://cddis.nasa.gov/pub/gps/products/ionex/}} to obtain $\rm RM_{Iono}$ for each epoch. The corrected RM, $\rm RM_{\rm PSR}$, is then obtained by subtracting $\rm RM_{iono}$ from $\rm RM_{meas}$. In general, we find our DM precision to be better than a few $ \times 0.01 \rm pc ~cm^{-3}$ except for PSR J1811$-$1736, whose significant scattering tail makes us less sensitive to precisely measure its DM and RM values. For those pulsars, where previous RM
measurements are available (see Table~\ref{tab:rvm}), we find that our measurements to be consistent 
with deviations between catalogue value and our measurement of  $\lesssim4\sigma$.

\subsection{Polarisation properties}
\label{sec:rvm}

Inspecting the polarisation properties of the obtained profiles can reveal potential calibration problems that would negatively affect the timing precision and often introduce systematics. The profiles shown here will therefore also serve as a reference to compare with during continuing timing observations. However, as we will demonstrate, they can also be helpful to achieve our science goals.

Overall, all pulsars show only a modest degree of polarisation. The linear polarisation is much lower than seen, for instance, in young pulsars (see e.g.~\citealt{kj05}). The exception is PSR J1157$-$5112, while PSR J1603$-$7202 stands out for having an unusually large degree of circular polarisation in its trailing component.

\subsubsection{Flat Position Angle Swings}

\begin{figure*}
    \centering
    \includegraphics[width=0.9\textwidth]{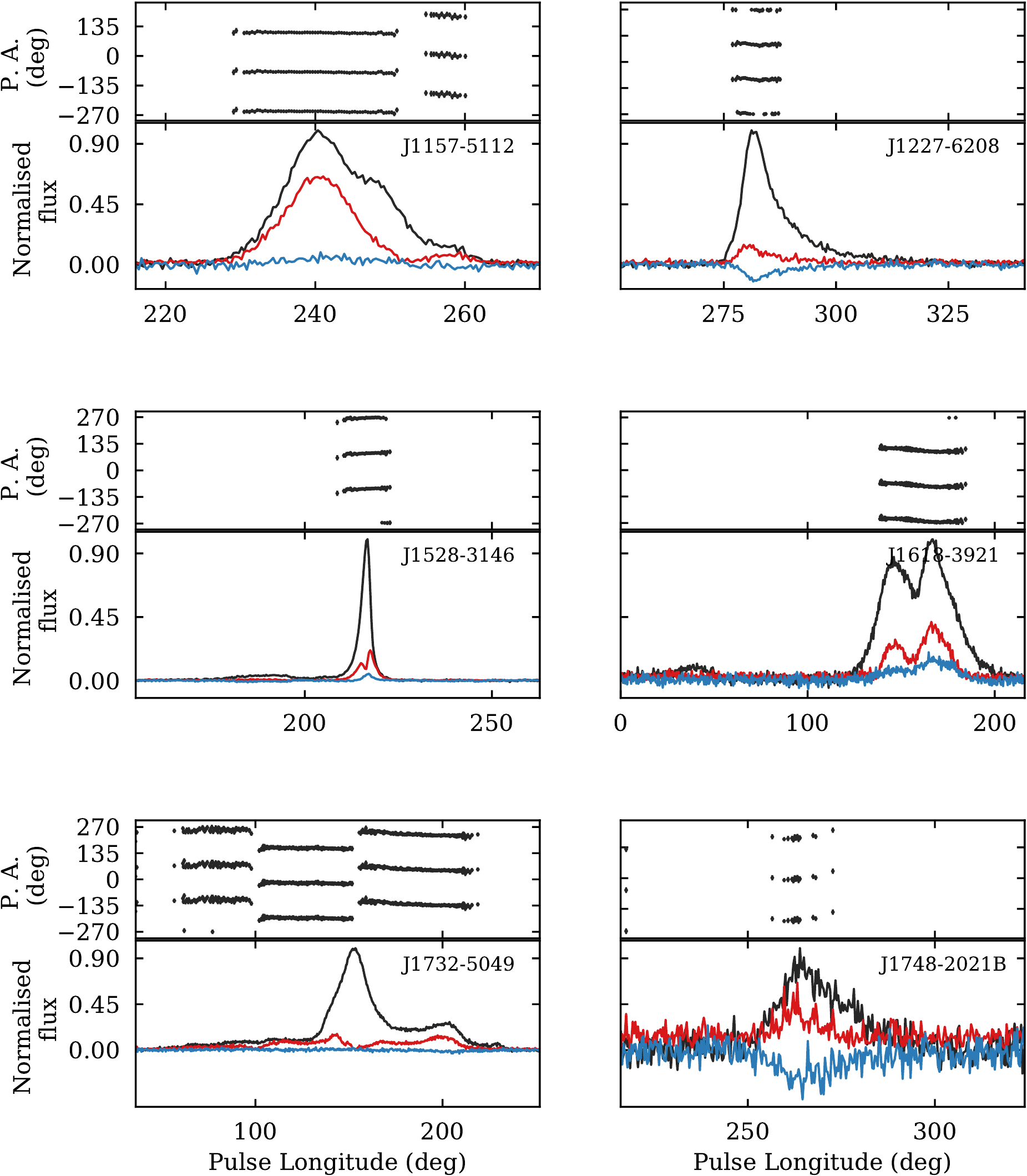}
    \caption{ {\color{black} Pulse Profiles for the relativistic binary pulsars with flat P.A. swings. See text for details. The data properties and panel descriptions are the same as Fig. \ref{fig:complexPAa}. }}
    \label{fig:flatPAa}
\end{figure*}

\begin{figure}
    \centering
    \includegraphics[width=0.45\textwidth]{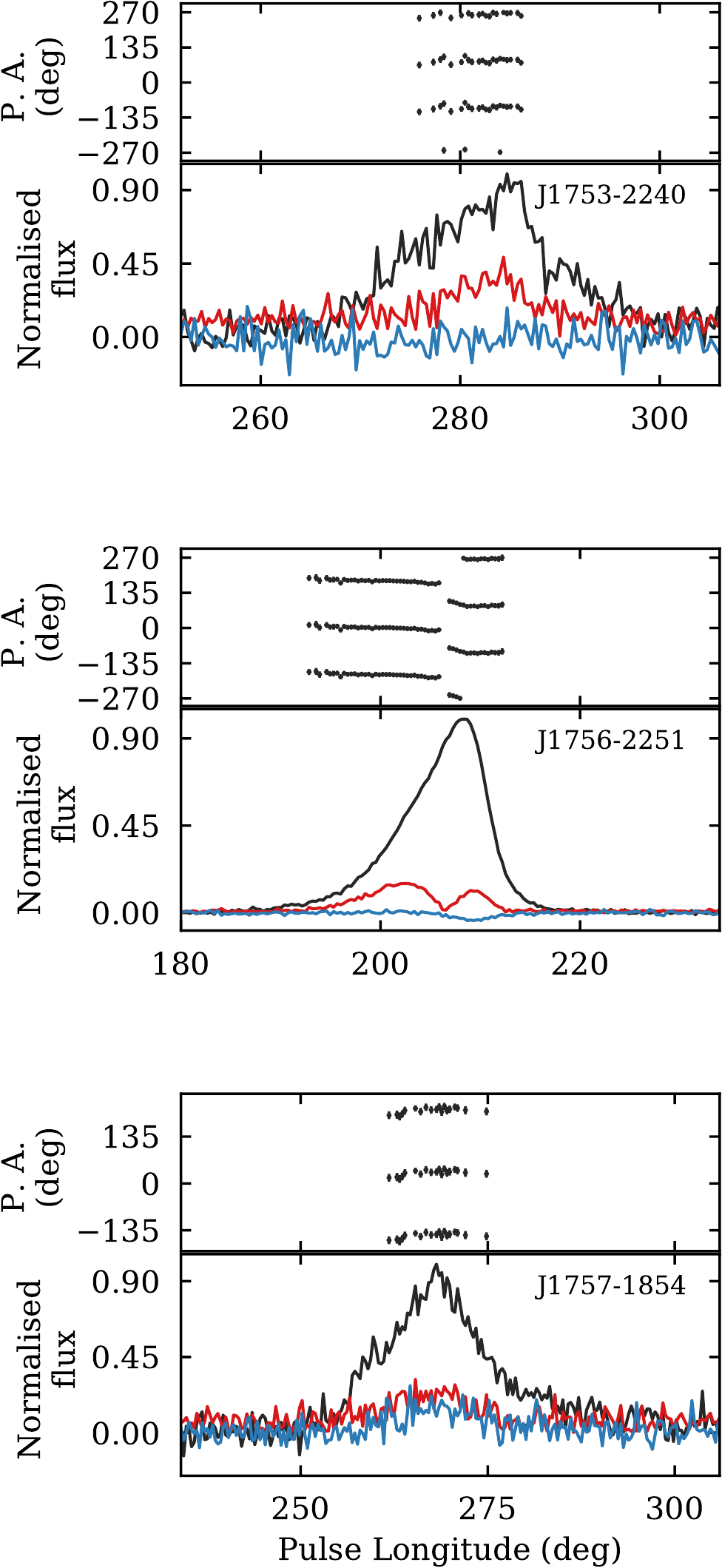}
    \caption{ {\color{black} Fig.~\ref{fig:flatPAa} continued.}}
    \label{fig:flatPAb}
\end{figure}

The large degree of linear polarsiation observed in PSR J1157$-$5112 is accompanied by a very flat P.A. curve. This, however, is not unusual for recycled pulsars, as shown already in early studies \citep{XilourisEtAl1998,stc99}. Here, these flat 
P.A. angles can be  clearly seen in more than a third of the 24 sources presented (see Figures~\ref{fig:flatPAa} \& \ref{fig:flatPAb} and Table~\ref{tab:flatPAs}).
We define sources as having ``flat'' P.A.s where we can measure the P.A.s for a sufficiently large range of pulse longitudes, while the P.A. behaviour is neither complex  (c.f.~Figs.~\ref{fig:complexPAa} \& \ref{fig:complexPAb}) 
nor similar to expectations from the  
Rotating Vector Model (RVM, \citealt{Radhakrishnan&Cooke})
 (c.f.~Fig.~\ref{fig:rvmPA}).
Some pulsars in this sub-set obviously show orthogonally polarised modes (e.g.~PSRs J1157$-$5112 and J1732$-$5049), which we account for before making the judgement. 

A flat P.A. curve may be caused by scattering, which does not only convolve the profile with an exponential
tail, but which also tends to flatten out the P.A.'s. This appears to be the case for PSRs 1227$-$6208. For other pulsars, if
one were to interpret those flat  P.A.s geometrically in the RVM framework  (see Sec.~\ref{sec:rvmfits}), one would assume an aligned geometry (in extended P.A. ranges) or in general, a grazing of the pulsar beam, far away from the magnetic pole. As we argue later, we believe that this is the case for PSR J1141$-$6545.

\begin{figure}
    \centering
    \includegraphics[width=0.4\textwidth]{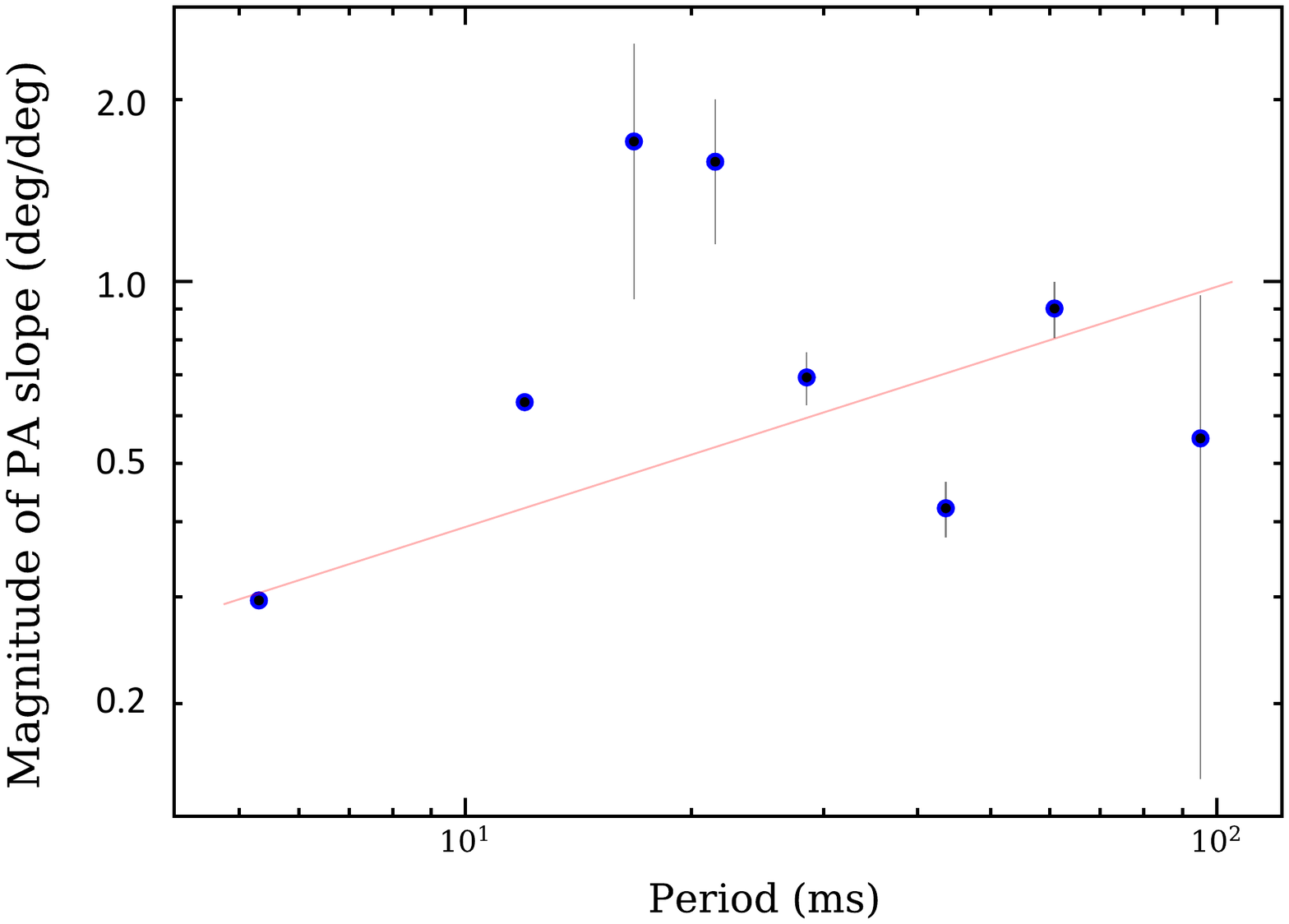}
    \caption{Magnitude of the slopes measured from P.A. swings as listed in Table~\ref{tab:flatPAs} as a function of pulse period. See text 
    for details.}
    \label{fig:slopes}
\end{figure}

Generally, fitting an RVM to the observed flat P.A. curves has little success and will not meaningfully constrain the viewing geometry. Alternatively, unusually flat P.A. swings may also be caused by
caustic emission \citep{man05,rmh10} or by
propagation effects in the magnetospheric plasma \citep{bp12,hbp17}. Such latter effects may depend on plasma densities and hence, potentially, on the 
magnetic field or spin and spin-down parameters.
While we will study this elsewhere, it is interesting to simply measure the slope of the position angle swing for those pulsars. 
Excluding PSRs 1227$-$6208 (scattered) and 1141$-$6534 (grazing cone), we
present our measurements (obtained via a straight-line fit to all P.A. values shown for each pulsar in Figures~\ref{fig:flatPAa} \& \ref{fig:flatPAb})
in Figure~\ref{fig:slopes}, where we show the magnitude of the slope, $|d{\rm P.A.}/d\Phi|$, as a function of pulse period. 
It is notable that a formal fit  reveals a weak dependence on the period as $|d{\rm P.A.}/d\Phi| \propto P^{0.40\pm0.02}$, 
suggesting that pulsars with smaller periods tend to show a shallower slope. Whether this
dependence is significant and confirmed with additional data remains to be seen.
We will defer the answer of this question to a more detailed and larger study in 
a later publication where we cannot only increase the sample size by adding
non-RelBin sources, but where we can also benefit from longer observing spans than available here.
Longer integration time may reveal additional P.A. values that may deviate from a flat P.A. swing observed
here. A larger sample can also look at possible physical origins, should this trend be confirmed, e.g. the possible dependence on the amount of accreted matter and other source-specific parameters. This is beyond the scope
of this paper.

\begin{table}
    \centering
        \caption{List of pulsars in our sample that show a flat P.A. swing. We list the measured slope, its uncertainty and
        the reduced-$\chi^2$ value to indicate how well the data can be described by a simple straight P.A. model. }
    \begin{tabular}{lcccc}
    \hline
    \hline
        PSR &  Period (ms) & P.A. Slope (deg/deg) & $\chi^2_{\rm
        red}$ \\
        \hline
        J1157$-$5112  & 43.59 & $-$0.42(4) &  0.9 \\
        J1528$-$3146  &  60.82 & +0.9(1) &  1.4 \\
        J1618$-$3921  &  11.99 &  $-$0.63(2) & 2.3 \\    
        J1732$-$5049  &  5.31 &  $-$0.296(4) & 2.2 \\
        J1748$-$2021B  &  16.76 &  +1.7(8) &  0.7 \\
        J1753$-$2240  &  95.14 &  +0.5(5) &  1.6 \\
        J1756$-$2251  &  28.48 &  $-$0.69(7) & 2.4  \\
        J1757$-$1854 & 21.50 &  +1.6(4) & 0.9 \\
         \hline
    \end{tabular}
    \label{tab:flatPAs}
\end{table}

With the currently available data set, for all cases shown in Table~\ref{tab:flatPAs} the magnitude of the slope is less than 2 deg/deg which is extremely flat, sometimes measured even over a wide range of longitudes. 
This is indeed difficult to explain in a geometrical model, although  we cannot rule out that some P.A. swings are the result of extreme aberration effects (which could also depend on period) or can be simply explained by sharp unresolved 180 degree swings for a central cut, both of which would still indicate a potentially valid geometrical interpretation of the P.A. swings.
Also, as discussed, flat PAs may still be representing the geometry of a grazing beam, as we believe is the case for PSR J1141$-$6545, a non-recycled pulsar,  based on additional information available to us via the observed relativistic spin-precession \citep{VenkatramanKrishnanEtAl2019}. We discuss this further in Section~\ref{sec:1141}, but we note here that if we were to include PSR J1141$-$6545 in Figure~\ref{fig:slopes}, it would continue the general trend with a slope of 3.82(5) deg/deg and a spin period of 394 ms. This would result in a slightly steeper power law index of $+0.59(1)$.

\subsubsection{Orbital inclination angle information from RVM}
\label{sec:rvmfits}

\begin{figure*}
    \centering
    \includegraphics[width=0.9\textwidth]{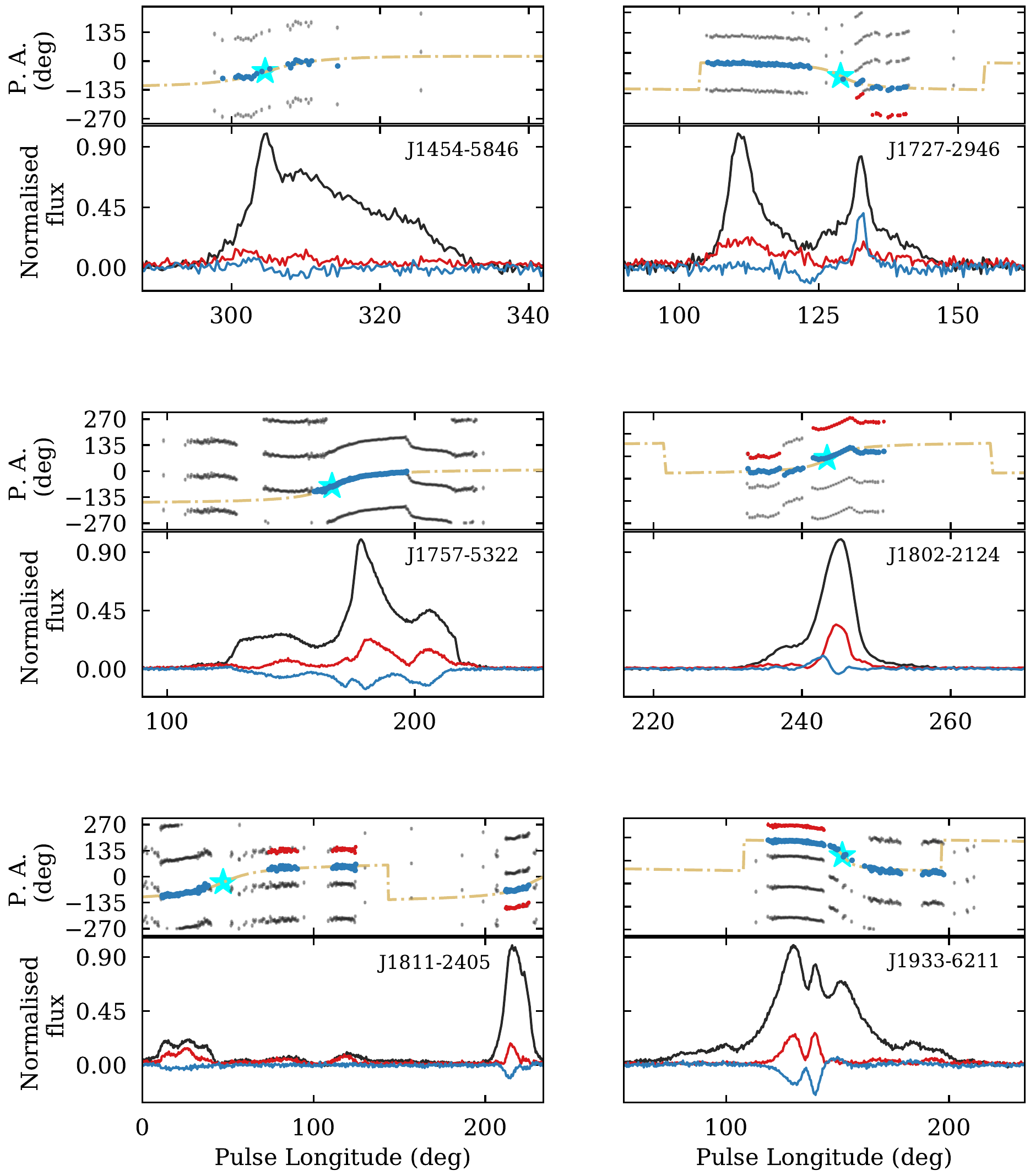}
    
    \caption{ { \color{black} Pulse profiles for relativistic binary pulsars, where we try to model the P.A. with a Rotating Vector Model (RVM) . The data properties and panel descriptions are the same as Fig. \ref{fig:complexPAa}. The sub panels showing the P.A.s also show the best-fit rotating vector model as a dash-dotted brown line. The black points in the panel are the original P.A. points - now made semi-transparent for clarity. The blue ones are the points that are considered for the RVM fitting. The red points are the original locations of some P.A. points that had to be shifted by 90 degrees to account for an orthogonal polarised mode transition, or shifted by 180 degrees for fitting convenience. The cyan star is the best-fit position of ($\phi_0$, 
    P.A.$_0$) - see text for more details.}}
    \label{fig:rvmPA}
\end{figure*} 

For a number of pulsars with a noticeable P.A. swing, combined with a sufficiently large number of well defined P.A. values, we have attempted to apply the RVM because recent results of the relativistic binary PSR J1906$+$0746 have indeed confirmed unequivocally that the P.A. swing has a geometrical origin according to the RVM, at least for some sources (\citealt{desvignes2019}).

\begin{figure*}
    \centering
    \includegraphics[width=0.9\textwidth]{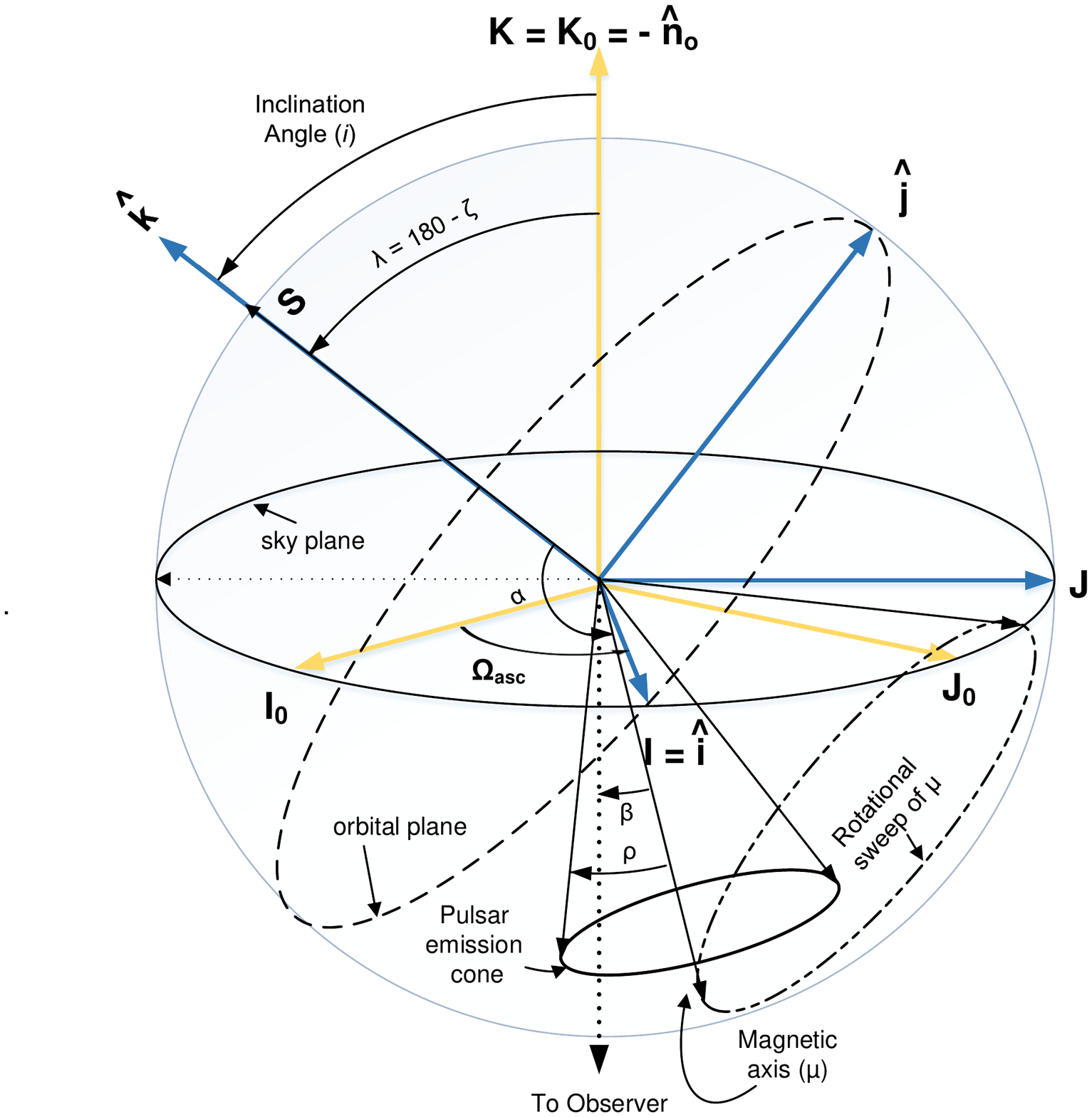}
    \caption{Definition of angles relevant for the viewing geometry of binary pulsars in a coordinate system defined by Damour \& Taylor (1992).
    The plane defined by {\color{black} $\hat{\mathbf i}$ and $\hat{\mathbf j}$} form the orbital plane which is inclined at an angle $i$ to the sky plane and rotated in azimuth by the longitude of the ascending node ($\Omega_{\rm asc}$). {\color{black} The sky plane is the plane perpendicular to the line of sight vector, $\hat{n_0} = - \mathbf K$, defined from the pulsar to the observer. } The spin angular momentum vector of the pulsar is given by \textbf{S}, here seen to be along the same direction of the orbital angular momentum vector, \textbf{K}. This is true for some binaries where the pulsar has been  fully recycled by a phase of mass accretion from the companion during its evolutionary history. For the cases where this is not true, please see Damour \& Taylor (1992) and Venkatraman~Krishnan et al. (2019). The magnetic axis of the pulsar ($\upmu$) is inclined from the spin axis by an angle $\alpha$. The radio emission cone subtends an angle $\rho$ from $\upmu$. The closest approach of the observer's line-of-sight to $\upmu$ marks the impact parameter $\beta$, which relates to the total viewing angle as $\zeta = \alpha + \beta$. The angle subtended by \textbf{S} and \textbf{K} is $\lambda$, which for pulsars with no spin-orbit misalignment follows $\lambda = 180 - \zeta = i$. Note that the polarisation angle $\Psi_0$ is defined as the angle subtended by the projection of \textbf{S} on the sky plane from \textbf{I}$_0$ in the clockwise direction as viewed from the \textbf{+K} direction, which is the opposite sense to the observer's convention - see text. }
    \label{fig:geometry}
\end{figure*}

The RVM by \cite{Radhakrishnan&Cooke} describes the position angle values,
as a function of the magnetic inclination angle $\alpha$, the viewing angle $\zeta$ and the pulse phase, $\phi$. We show its modified form as presented in \citet{jk19}:
\begin{equation}
\label{eqn:rvm}
{\rm P.A.} = {\rm P.A.}_{0} +
{\rm arctan} \left( \frac{{\rm sin}\alpha
\, {\rm sin}(\phi - \phi_0 - \Delta)}{{\rm sin}\zeta
\, {\rm cos}\alpha - {\rm cos}\zeta
\, {\rm sin}\alpha \, {\rm cos}(\phi - \phi_0 - \Delta)} \right)
\end{equation}
Here, $\phi_0$ is the pulse longitude at which PA=PA$_{0}$ and $\zeta=\alpha+\beta$ (see Fig.~\ref{fig:geometry}). The additional $\Delta$ term is present to deal with cases in which the emission heights are different between the main pulse and a potentially observed interpulse. 

\begin{figure}
    \centering
    \includegraphics[width=0.45\textwidth]{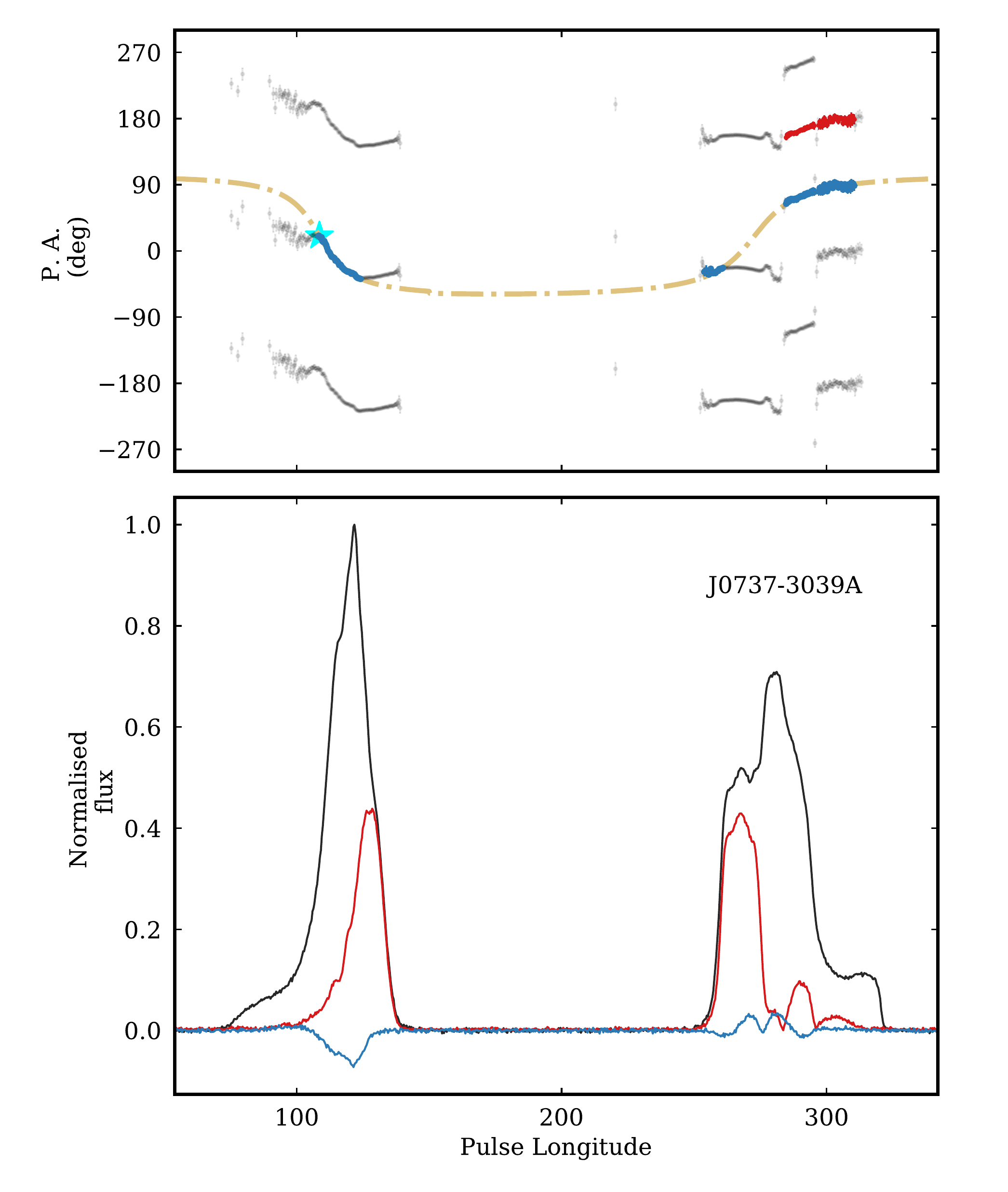}
    \caption{Fit of the modified RVM model given in equation \ref{eqn:rvm} to the P.A. of PSR~J0737$-$3039A obtained from a 2-hour observation of the pulsar. The P.A. points are duplicated for clarity. The colors and symbols are same as Figure \ref{fig:rvmPA}}
    \label{fig:0737PA}
\end{figure}

We present the results of our RVM fits in the form of the determined magnetic inclination angle $\alpha$ and the viewing angle $\zeta$ in Table \ref{tab:rvm} and Figs.~\ref{fig:rvmPA} and
\ref{fig:0737PA}.
We note that we have not been successful in modeling the P.A. swings using the RVM for PSRs  J1017$-$7156, J1435$-$6100, J1603$-$7202,  J1930$-$1852, J2129$-$5712 or J2222$-$0137 (see Figs.~\ref{fig:complexPAa} \& \ref{fig:complexPAb}).
This lack of success is not surprising and adds to the general notion mentioned before. For PSR J0737-3039A we present additional data, which we discuss in detail in Section~\ref{sec:dpsr}.

For the other sources,
in order to obtain successful RVM fits, we have used the methods described in \cite{jk19}. Using the arguments outlined there, before the fits, 
 we have sometimes introduced orthogonal jumps at certain pulse longitudes when a drop in linear polarisation suggested that this is possible. In a few cases, following again \cite{jk19}, we have ignored certain pulse phases where we considered it possible that unresolved and overlapping orthogonal modes may have led to intermediate PAs resulting from a mixture of modes. We indicate those choices in Figure \ref{fig:rvmPA}.
 
The geometry derived and presented in Table~\ref{tab:rvm} should therefore be taken with the usual caution. Nevertheless, in particular the viewing angle $\zeta$ of recycled pulsars is of interest. For fully recycled pulsars, from evolutionary arguments, we expect the spin vector of the pulsar to be aligned with the orbital momentum vector. For systems that show relativistic spin-precession, we may be able measure the angle between pulsar spin and orbital momentum vector, $\delta$, as for PSR J1906$+$0747 ($\delta=104\pm9$ deg, \citealt{desvignes2019});
 or for those DNS components that do not precess, we may be able to derive a stringent upper limit on the misalignment angle, as for PSR J0737$-$3039A (i.e.~$\delta<3$ deg (95\% cl.) \citealt{fsk+13}). In all these cases, we can  compare 
$\zeta$ with the orbital inclination angle, $i$. Hence, determining $\zeta$ via a successful RVM fit offers a way to determine $i$ independently of a Shapiro delay measurement, which only allows a measurement of  $\sin i$. Hence, apart from providing important information for tests of gravity or mass measurements (e.g., solving the mass function), RVM fitting may indeed also enable us to solve the corresponding $i$ or $180-i$ ambiguity of a Shapiro delay measurement.  

Comparing the value of $\zeta$ with the orbital inclination angle requires caution. We stress that depending on the convention used for the measurement of the position angle and the applied RVM equation, one needs to identify $\zeta$ either with $i$ or $180-i$. We will demonstrate the power of this additional information with two specific examples, namely the Double Pulsar and PSR 1811$-$2405, and we elaborate on the situation to provide useful guidance for future studies.

The position angles in Figs.~\ref{fig:complexPAa} - \ref{fig:flatPAb}, \ref{fig:rvmPA}, \ref{fig:0737PA} and \ref{fig:J1141-6545}
are measured in the so-called ``observer's convention'' with the angles increasing counter-clockwise on the sky.
This convention is adopted by the \cite{IAU}. The standard pulsar software \textsc{psrchive} used here follows this convention, also known as the PSR/IEEE
convention \citep{vmjr10}. As pointed out by \cite{DamourTaylor1992}, this convention differs from the convention where the position angle increases clockwise on the sky. This latter convention was used in the definition of the RVM \citep{Radhakrishnan&Cooke}, so that it is also referred to as the ``RVM convention''. \cite{EverettAndWeisberg2001} clearly pointed out the consequences for the angles derived from a RVM, namely the magnetic inclination angle $\alpha$ and the viewing angle, $\zeta$, or the the impact angle $\beta$, where $\zeta=\alpha+\beta$ (see Fig.~\ref{fig:geometry}). Importantly, the definition of $\zeta$ in the RVM binds it to the definition of $i$. Using a definition of orbital geometry as shown in Fig.~\ref{fig:geometry}, which is derived from \cite{DamourTaylor1992} (see also~\citealt{kw09}), and, crucially,
is in contrast to the implementation in the timing software
\textsc{Tempo}\footnote{\url{http://ascl.net/1509.002}}  and \textsc{Tempo2}
\footnote{\url{http://ascl.net/1210.015}},
we identify
\begin{equation}
    \zeta = 180 - i,
\end{equation}
when pulsar spin axis and the orbital angular momentum vector are aligned. We call this definition also the ``DT92'' convention and refer to the definition of the corresponding position angles as the ``RVM/DT92'' convention.

As \cite{EverettAndWeisberg2001} before, we strongly recommend to refer to the pulsar geometry angles in the RVM/DT92 convention. This can be easily identified. The impact angle is {\em positive}, $\beta>0$,  i.e.~$\zeta>\alpha$, 
when the slope in the position angles derived by the observer's PSR/IEEE convention (as for \textsc{psrchive}) and
measured at its steepest gradient (at the centroid or fiducial plane) is {\em negative}, and vice versa. 

As Eqn.~\ref{eqn:rvm} is written in the RVM/DT92 convention, fitting it to P.A.s measured in the PSR/IEEE convention (as in our figures)
requires an important prior step. Either one inverts the P.A. values by multiplying them by $-1$, or one modifies Eqn.~\ref{eqn:rvm} such that one fits for the arguments $(\phi_0-\phi)$ rather than $(\phi-\phi_0)$, where we set $\Delta=0$ for clarity. In both cases, this modification effectively transfers the P.A. from counter-clockwise to clock-wise rotation (RVM/DT92 convention) before deriving the angles. For this work,  consistent also with \cite{jk19}, we implemented the latter. The resulting angles listed in Table~\ref{tab:rvm} are therefore given in RVM/DT92 convention.

In general, due to the co-variances in $\alpha$ and $\zeta$ caused by the structure of Eqn.~\ref{eqn:rvm}, the uncertainties of the angles derived from RVM fits are usually very large if the range of fitted pulse longitudes is limited (see e.g.~discussion by \citealt{EverettAndWeisberg2001} and \citealt{Lorimer&Kramer2005}). 
However, when the pulses are wide, and especially when interpulse emission is seen, the precision in the derived angles is much improved. This can also be seen for the results shown in Table~\ref{tab:rvm}. The angles of PSRs J0737$-$3039A and J1811$-$2405 have a good formal precision and
are discussed further below.
 We note that the quoted values
and their uncertainties correspond to the median and the 16 and 84 percentiles of the posterior distribution, respectively. If larger confidence limits are required, the uncertainties scale accordingly. We also point out that the results are, of course, also affected by our informed choice of P.A. values that are included in the fit. This can be demonstrated in the case of PSR J1757$-$5322. The result shown in Table~\ref{tab:rvm} makes use of only the central P.A. values (see Figure~\ref{fig:rvmPA}). These still span 40 deg in longitude and hence already constrain the geometry well enough to yield a relatively small error of 7 deg for both $\alpha$ and $\zeta$. The implied orbital inclination angle of $i = 94\pm7$ deg would suggest a Shapiro delay that probably should have been measured already (cf.~Section~\ref{sec:1757}). However, for the 95\% uncertainty limits, $\zeta$ is still consistent with a range from 71 to 99 deg, and the orbital inclination correspondingly. Including more P.A. values from the wings of the profile, pushes the solution to smaller $\zeta$ (and $\alpha$) values, which is a systematic uncertainty that is not yet reflected in the quoted statistical uncertainties. To gauge the impact on the overall result, we mark the P.A. values included in each fit in Fig.~\ref{fig:rvmPA}, so that readers can form their own opinion about the reliability of the results.

We emphasize that one way of constraining the uncertainties is by using an informed non-uniform prior on $\zeta$. With $\zeta=180-i$ for aligned spin and orbital momentum vectors, we can take into account that a face-on orbit (small $i$) is less likely be found than an inclined orbit, by using a uniform prior on $\cos \zeta$.  Similarly, if we have good constraints on the pulsar and/or companion mass (e.g. from optical observations of the companion), one can also use the mass function to construct a prior on $i$ and, hence, $\zeta$. Implementing this had little impact on our results compared to those obtained with a uniform prior, but there are clearly cases, where such strategies will be useful 
as we demonstrate for PSR J0737-3039A in Section~\ref{sec:dpsr}.
Similarly, one may also adopt non-uniform priors for $\alpha$, for instance in studies that include a possible variation of the magnetic inclination angle on long timescales. 
The results shown in Table~\ref{tab:rvm} were obtained with uniform priors on the angles, 
except in the case of PSR J0737-3039A.

In summary, we clearly recommend treating the results of RVM fits, especially to mildly or fully recycled pulsars with care. On the other hand, we do have a number of pulsars the RVM fits are compelling: 
We discuss PSR J0737-3039A using a polarisation profile based on a longer
observation (see Figure~\ref{fig:0737PA}) in more detail in Section~\ref{sec:dpsr}.
We further discuss PSR J1811$-$2405 in Section~ \ref{sec:1811} to demonstrate the power of this method.
We conclude that the results may at least give a useful general impression about the 
overall geometry with its implication for the $\sin i$ ambiguity or estimating whether a Shapiro delay measurement may be possible at all. Indeed,  the usage of the ``pulse structure information'' -- given by the profile and its polarisation properties -- can be very powerful.
This information complementary to the timing, cannot only help to determine orbital inclination angles for mass determinations, but especially also for tests of gravity. This is demonstrated here, 
consistent with the recent work by \cite{desvignes2019} or earlier by \cite{Stairs2004}.

\begin{table}
    \centering
    \caption{Viewing geometry as derived from fits of the RVM to a subset of the pulsars. The magnetic inclination angle $\alpha$ and the viewing angle $\zeta$ are given. Their values and uncertainties quoted correspond to 
    the median and the 16 and 84 percentiles of the posterior distribution, respectively. The last column lists the orbital inclination angle, $i$, as implied from the obtained $\zeta$ value (see text for details). }
    \begin{tabular}{lccc}
    \hline
    \hline
    PSR & $\alpha$ (deg) & $\zeta$ (deg) & Implied $i$ (deg)\\
    \hline
    J0737$-$3039A & 79.1(1) & 88.4(1)$^*$ & 91.6(1) \\
    J1454$-$5846 & 133 (34) &  128 (33) & 52 (33) \\
    J1727$-$2946 & 121 (7) & 123 (7) & 57 (7) \\
    J1757$-$5322 & 96 (7) & 86 (7) & 94 (7)\\
    J1802$-$2124 & 57 (19) & 55 (19) & 125 (19) \\
    J1811$-$2405 & 91.5 (2)  & 76.0 (3) & 104.0 (3) \\
    J1933$-$6211 & 141(4) & 144 (4) & 36 (4)\\
         \hline
    \end{tabular}
    
    	 \begin{tablenotes}
	 	 \small
	 	 \item $^*$ The priors for $\zeta$ were restricted using existing timing\\ constraints on the inclination angle. 

	    \end{tablenotes}
    \label{tab:rvm}
\end{table}

\begin{figure*}
    \centering
    \includegraphics[scale=0.85 ]{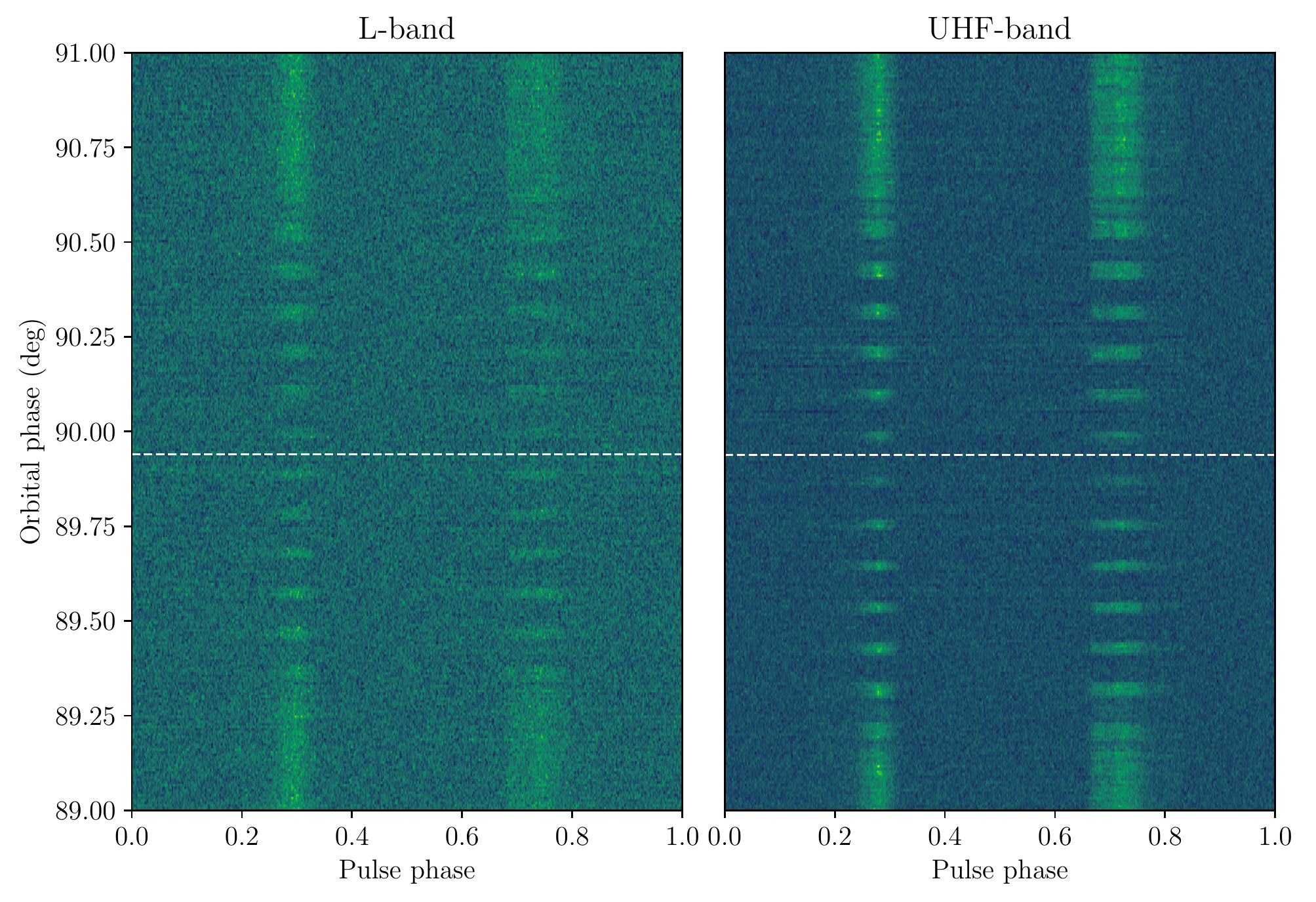}
    \caption{\color{black} A plot of the pulsed intensity modulation of J0737$-$3039A at L- and UHF-bands during superior conjunction when it is eclipsed by the magnetosphere of its companion. The colour scale (from blue to green) denotes the modulations in the total intensity of the pulse. In both bands, each integration is a sum of 8 pulses ($\sim 180$ ms). It can be clearly seen that the modulation is at the spin period of the companion, which is $\sim$2.7 seconds. Note that due to the effect of retardation, the (apparent) superior conjunction of the pulsar is slightly shifted from 90 degrees of orbital phase, and is denoted by the dotted white lines.   }
    \label{fig:dpsr_eclipse}
\end{figure*}

\subsection{Timing Performance and Science Goals}
\label{sec:predictions}

Our science goals and measurement strategy as outlined in Section~\ref{sec:RelBinProgramme} usually require an extended timing baseline, which for all sources selected for RelBin at this point 
can be provided from previous observations with other telescopes. In order to reliably predict the expected measurement precision for the PK parameters and masses that we want to measure within RelBin, dedicated studies like those by \cite{huhu2020} need to be conducted.
However, as indicated in Section~\ref{sec:RelBinProgramme},
we can already gauge the potential of RelBin by comparing the timing precision that we obtained with MeerKAT thus far, with that presented for each pulsar in prior literature. We emphasize that
this is a conservative estimate, since our values listed in Table~\ref{tab:expectations}
are obtained for a standard observing and processing set-up. In a final
analysis, the choice of receivers (``UHF'', ``L-Band'' or ``S-Band'') and analysis pipeline
(e.g.~frequency-evolving templates for our wide-band data) will be optimized for every source
individually. Hence, the already visible, often large improvement in the timing precision is impressive and fills us with great confidence that we can achieve our objectives. We comment on selected individual cases below and also refer to the PTA programme (Spiewak et al. in prep.) and future publications dedicated to the various sources for more details.

\section{Initial Results for Selected Sources}

\label{sec:specific_sources}

\subsection{Sources for Tests of Gravity}

\begin{figure}
    \centering
    \includegraphics[width=0.5\textwidth, trim={20 0 0 0}]{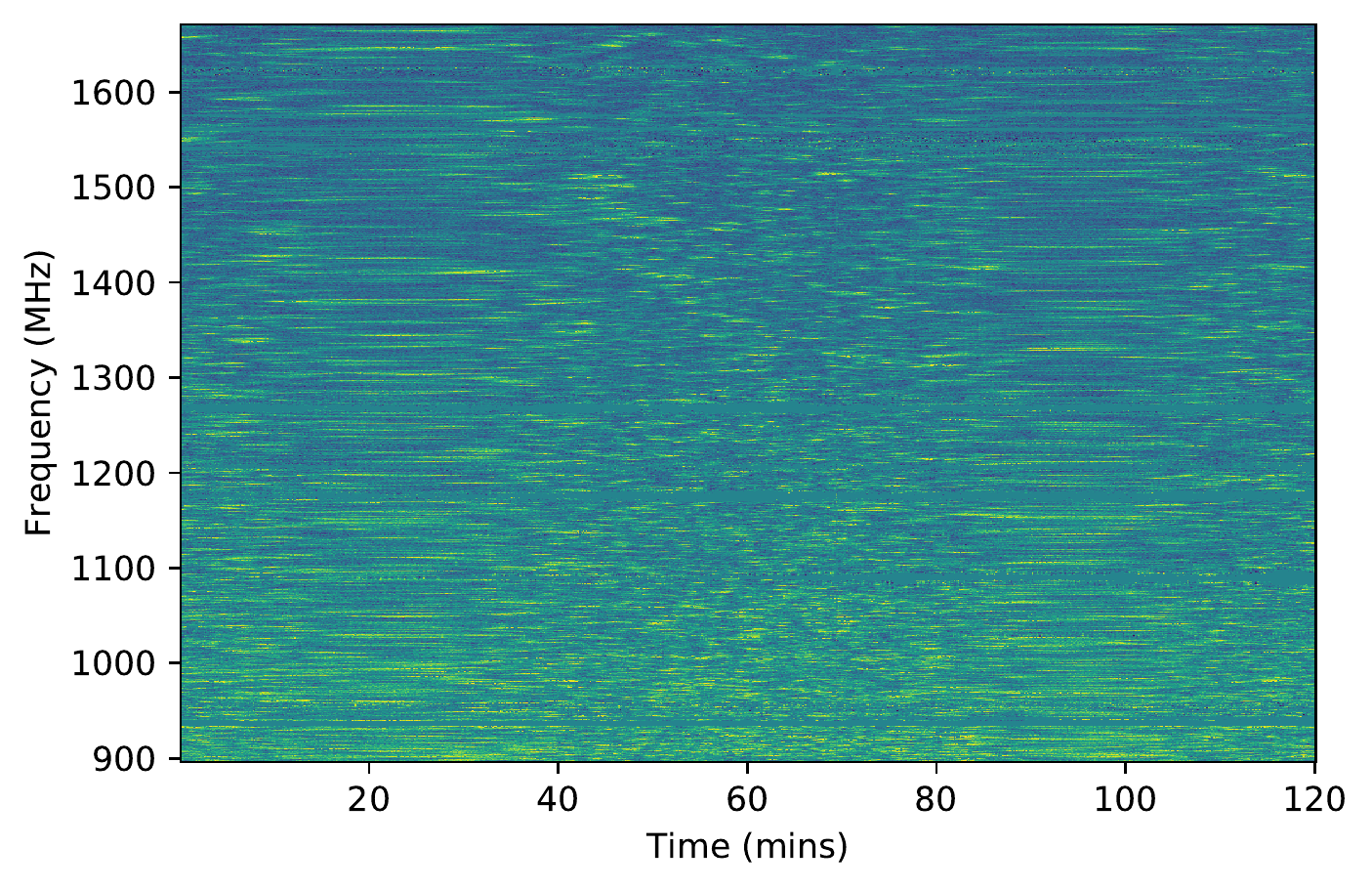}
    \caption{Dynamic Spectrum of the Double Pulsar, obtained from a 2-hr observation of the pulsar. The stretching and squeezing of the blobs of power are caused due to the pulsar's orbital motion.}
    \label{fig:dpsr_dynspec}
\end{figure}

\subsubsection{The Double Pulsar - PSR J0737$-$3039A/B}
\label{sec:dpsr}

The Double Pulsar is a unique double neutron star system where both component neutron stars in the system are active radio pulsars. Due to relativistic spin-precession, the radio pulse of the companion (J0737$-$3039B) precessed out of our line of sight in 2008 and is no longer visible until it returns. Nevertheless, timing both pulsars has facilitated the measurement of a variety of Keplerian and relativistic effects, which in conjunction have resulted in six tests of gravity \citep[cf.][]{KramerEtAl2006,BretonEtAl2008,Kramer&Wex2009}. The Double Pulsar has historically been timed with a variety of telescopes including the GBT, Lovell, Effelsberg, Nan\c{c}ay and Parkes telescopes. 
Our observations with MeerKAT at L-band are twice as sensitive as those from GBT or Effelsberg (owing to the very low elevation angles that those telescopes needed to observe this source). This increase in sensitivity has opened up the possibility of the detection of several higher order post-Keplerian effects and to study its eclipses in high temporal resolution. 

While the pulsar was timed for a full orbit roughly once every 3-4 weeks with the L-band system for the last 18 months, we have recently started observing at the UHF-band, and 
we find that the timing precision is $\sim 1.6~\times$ better at UHF than at L-band. 
The improved timing precision compared to less-sensitive observations and between L-Band and UHF that has been achieved  already, suggests that Double Pulsar timing will not be limited by pulse jitter, at least until the SKA comes online \citep{huhu2020}. With MeerKAT 
we will continue to time the pulsar at both frequency bands, as significant DM variations are observed (Kramer et al., in prep.). This dual-frequency approach allows us to better connect our MeerKAT data with existing data sets from other telescopes.
  
The polarisation profile shown in Fig.~\ref{fig:complexPAa} confirms the polarisation data presented by \cite{KramerStairs2008} and is specifically consistent in the degree of polarisation, the sense of circular polarisation and the direction of the P.A. swing. We note that this is in contrast to earlier polarisation data published by \cite{DemorestEtAl2004} or  \cite{HotanEtAl2005b}. In order to determine the geometry, \cite{DemorestEtAl2004} applied RVM fits to their data with the conclusion that they favoured an aligned (small $\alpha$) geometry.  
 In contrast, \cite{gkj+13} attempted to fit a standard RVM to the polarisaton data of \cite{KramerStairs2008} and derived values for $\alpha$ and $\zeta$ close to 90 deg. An orthogonal geometry was also derived by \cite{fsk+13}, based on considerations of the profile stability in the possible context of spin-precession.
We decided to improve further on the pulse profile shown in Fig.~\ref{fig:complexPAa} with additional observing 
time. The resulting profile is shown, also with more detail, in Fig.~\ref{fig:0737PA}. Fitting  
Eqn.~\ref{eqn:rvm} to these data results in the geometry presented in Table~\ref{tab:rvm}.
We confirm that the pulsar is in fact an orthogonal rotator,  i.e.~we see the emission from opposite magnetic poles.
As we explain below, in order to derive this result,
we used the information that is available on the orbital inclination angle, i.e.~we observe an edge-on orbit. The measurement of a Shapiro
delay gives a value of $i=88.7 (-0.8, +0.5)$ deg, or $i=91.3 (-0.5, +0.8)$ as measured by \cite{KramerEtAl2006}, whereas an extended data set suggests an angle somewhat closer to 90 deg (but consistent with the previous value, Kramer et al., in prep.) and by modelling of the relativistic spin-precession of pulsar B (J0737$-$3039B) \cite{BretonEtAl2008}.

The edge-on geometry is also confirmed by the eclipse of pulses from PSR J0737$-$3039A during superior conjunction \citep{LyneEtAl2004}. High-time resolution observations by the GBT revealed that the lightcurve of PSR J0737$-$3039A during each eclipse is modulated at B's rotation period \citep{McLaughlinEtAl2004}, a phenomenon that is consistent with synchrotron absorption by plasma confined within the dipole magnetosphere of pulsar B
\citep{Lyutikov+Thompson2005, BretonEtAl2012}. 
Matching changes in the eclipse light-curve over time with the expected geometry of the rotating dipole model from \citet{Lyutikov+Thompson2005} enabled a novel measurement of relativistic spin-precession of and a sixth independent test of gravity with the Double Pulsar \citep{BretonEtAl2008}. 

We can demonstrate the increased sensitivity of MeerKAT in Figure \ref{fig:dpsr_eclipse}, where the eclipse seen 
with the UHF-receiver has a S/N almost three times better than the eclipse measured with the MeerKAT L-band receiver. 
We note that the UHF-eclipse is also measured three time better than with the GBT at the same frequency range (due to larger sensitivity and less spill-over contribution to the system temperature).
Combined with the improved polarisation purity of MeerKAT, these observations will enable detailed studies of spectro-polarimetric variations in the pulses of PSR J0737$-$3039A due to propagation effects within B's magnetosphere at the ingress and egress of the eclipse, as proposed by  \cite{Lyutikov+Thompson2005} and \cite{BretonEtAl2012}, and tentatively seen with Parkes by  \cite{YuenEtAl2012}. In addition, advances in stochastic sampling algorithms and an augmented iteration of the methodology of \citet{BretonEtAl2008} will enable precision timing of PSR J0737$-$3039B through direct fitting of individual eclipse light-curves. This allows for a significantly improved measurement of its relativistic spin precession and the associated test of gravity.

Indeed, using simulations with a comparable S/N to the data shown in Figure 9, we are able to estimate the expected improvement in the measurement of B's precession rate,  $\Omega_{\rm B}$, when compared to the results of Breton et al. (2008). We find that we can obtain a measurement of  $\Omega_{\rm B}$ with comparable precision using only two independently measured eclipses separated by 1.5 years. By combining these eclipses with the results of Breton et al. (i.e. referring back to December, 2003), we can expect to measure  $\Omega_{\rm B}$ to about 1\% accuracy, thereby improving the precision of this gravity test by about an order of magnitude or more. These estimates are somewhat conservative, as we will have many more eclipses available from our monthly observations, in addition to current work on improving the eclipse analysis tools. 

We indicated that the RVM modelling presented in Section~\ref{sec:rvmfits} 
was done with an uniform prior on $\zeta$ between 
$(88.7-0.8-3.0)$ deg and $(91.3+0.8+3.0)$ deg.
 Extending the prior range by an additional 3 deg, beyond the $0.8$ deg uncertainty of the Shapiro
 delay measurement, takes account of the upper limit on the spin-misalignment angle  of $\delta<3$ (95 C.L.) by \cite{fsk+13}.
In the modelling, we find the flat sections of the P.A. in both main and interpulse difficult to describe by the RVM. Since they are preceding clear OPMs, and their flat structure does not fit a general RVM, we chose to ignore them, even though the linear polarsiation is highly significant here (see Figure~\ref{fig:0737PA}). With this caveat in mind, the obtained result is otherwise very convincing. However, we note that the large $\Delta=-15.7(3)$ deg value can be interpreted as a significant lower emission height for the interpulse, when compared to the main pulse.
In order to check the robustness of the result, we also modelled the two poles separately, with separate RVMs. Interestingly, the obtained $\alpha$ values, and even the P.A.$_0$ values (which is not necessarily expected) are perfectly consistent, not only with each other but also with the joint fit presented earlier, albeit with larger uncertainties as expected.

As a further check, we can see if the pulse widths are consistent with the obtained geometry. 
For both main and inter pulse, we measure a pulse width at a 10\%-intensity level of about $W\sim72$ deg
using a method described by \cite{kwj+94}.
We can use the following relationship 
\begin{equation}
\label{rho}
\cos\rho = \cos\alpha\,\, \cos \zeta\,\, +\,\,  \sin \alpha\,\, \sin\zeta\,\, \cos ( W/2)
\end{equation}
\citep{ggr84}  to  infer the beam radius, $\rho$. We obtain $\rho_{\rm MP}=37.2$ deg for the main pulse and $\rho_{\rm IP}=37.6$ deg for the interpulse. This can be compared to a value of $\rho$ that we expect if the pulsar follows a known $\rho = k \times P^{-0.5}$ scaling relationship, whereas $k$ ranges from
4.9 to 6.5 deg s$^{0.5}$, 
resulting in an uncertainty that is larger than that of the measured width
(see the discussion by \citealt{VenkatramanKrishnanEtAl2019}). Nevertheless, if
this can be applied here, we expect $\rho$ to be between 33 and 44 deg, which is in excellent agreement with our estimate derived from the RVM. It is also notable that the geometry derived from the RVM allows naturally for the poles to have the same beam radius and a resulting equal pulse width. This cannot be necessarily expected (from random combinations of $\alpha$ and $\zeta$) but agrees with the observations. 

Despite the prior for $\zeta$ ranging uniformly between $\sim85$ deg to $\sim95$ deg, the fit converges on $\zeta=88.4\pm0.1$ deg, clearly below 90 deg. This implies an inclination of $i=91.6(1) > 90$ deg. We note that this value is larger than an updated timing value reported later (Kramer et al.~in prep.), but most importantly it is inconsistent with the result by \cite{RickettEtAl2014} who used scintillation measurements to derive $i = 88.1(5) <90 $ deg .

Indeed, the Double Pulsar is among a small number of binary systems for which the system geometry can also be inferred from interstellar scintillation properties using a method that was pioneered by \cite{lyne1984}.
The dynamic spectrum in Figure \ref{fig:dpsr_dynspec} shows variation in pulsar A's flux due to interstellar scintillation. The scintillation timescale varies with orbital phase, owing to the changing transverse component of the pulsar's orbital velocity. Long term monitoring of this scintillation provides a way to uniquely determine parameters of the orbit including the inclination angle $i$ and longitude of ascending node $\Omega$ \citep{lyne1984,Ord2002_SV,RickettEtAl2014, Reardon+19}. As Figure \ref{fig:dpsr_dynspec} shows, we can clearly apply the method to the Double Pulsar, which will allow us to either confirm the earlier results by \cite{RickettEtAl2014} or decide in favour of the RVM estimate. We note that a geometrical model of scattering is required in order to correct for any significant time-variability in the spatial scale of the scintillation pattern \citep{Cordes+98}. Measurements of the scintillation over wide bandwidths, for example with near-simultaneous L-band and UHF observations, can be used to improve this model. If the geometry can be understood, modelling of the dynamic spectrum can also be used to estimate the pulsar distance and proper motion \citep{Reardon+19}.

\subsubsection{PSR J1141$-$6545}
\label{sec:1141}

\begin{figure}
    \centering
    \includegraphics[width=0.45\textwidth]{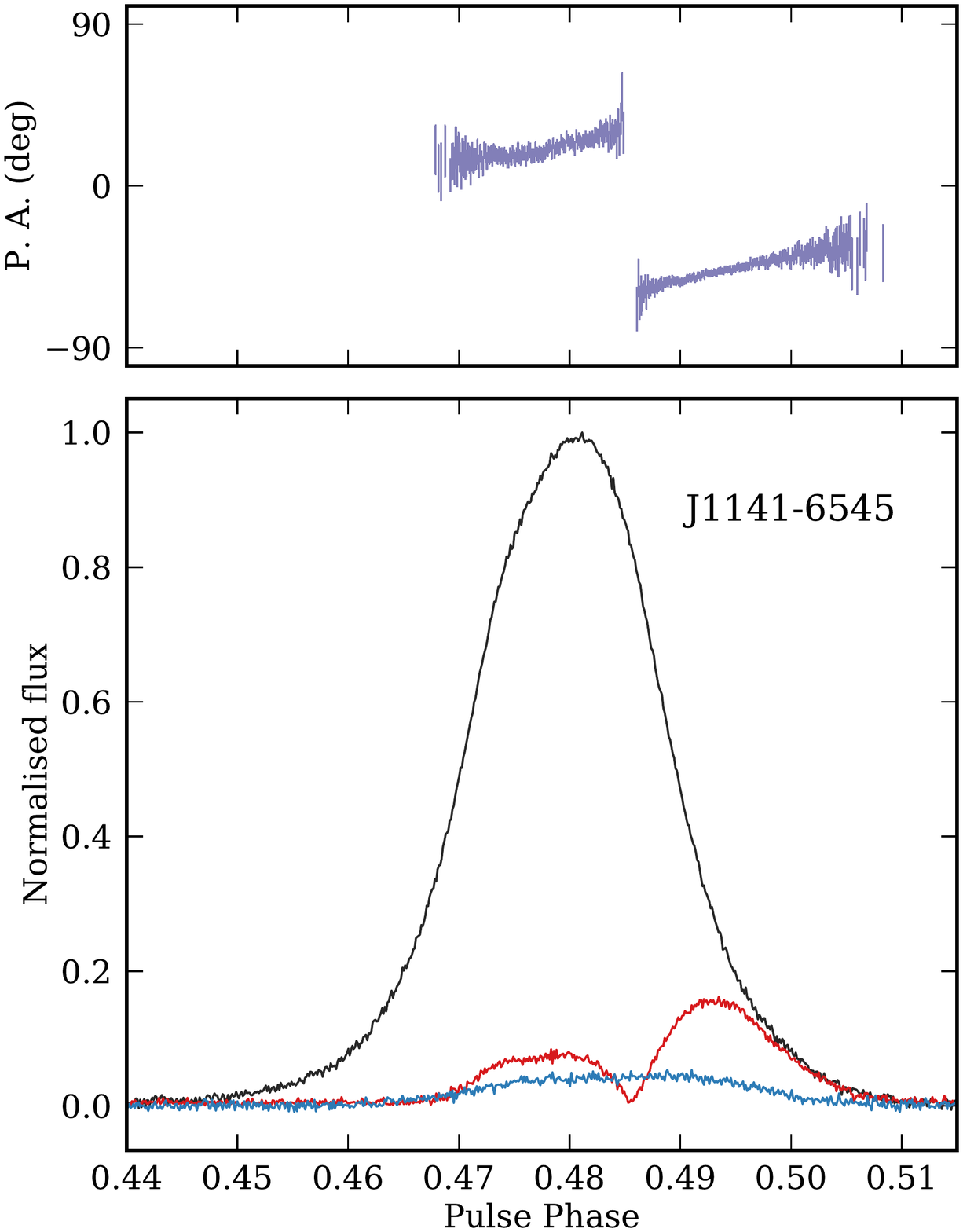}
    \caption{A high time-resolution plot of the polarised intensity profile of PSR J1141$-$6545, zoomed into the on-pulse region. The bottom panel shows the total intensity, linear and circular polarisation profiles plotted in black, red and blue respectively. The top panel shows the corresponding position angle swing of the linear polarisation. Our observations provide about 500 phase bins across the on-pulse region, clearly revealing a sudden jump in the P.A. swing, accompanied by a dip in the linear polarisation. Combined, these indicate an orthogonal polarised mode transition which was suspected but previously not confirmed \citep{ManchesterEtAl2010}. This crucial information will be highly beneficial for further modelling of the relativistic spin-precession of this pulsar.}
    \label{fig:J1141-6545}
\end{figure}

PSR J1141$-$6545 is a $394$ ms period pulsar in a $4.74$-hr orbit around a (CO or ONeMg) white dwarf companion. The long term timing of this pulsar with the Parkes and UTMOST radio telescopes have revealed a precession of the pulsar orbit due to classical and Lense-Thirring effects induced by the rapid rotation (period $\lesssim$ 200\,s) of the white dwarf companion \citep{VenkatramanKrishnanEtAl2020b}. This is the first known binary pulsar where the astrophysical interpretation of the orbital dynamics
requires contributions from relativistic spin-orbit coupling, known as the Lense-Thirring effect; this is non negligible fraction of the total observed spin-orbit coupling. Further timing of this pulsar will help to better constrain the spin period of the white dwarf \citep[cf.][]{Wex98,WexKopeikin1999}.

The pulsar also undergoes geodetic precession, leading to a secular change in its pulse shape. Geodetic precession of PSR J1141$-$6545 was initially analysed by \cite{HotanEtAl2005b}, while \cite{ManchesterEtAl2010} used a precessional RVM \citep{Kramer&Wex2009} to obtain estimates for the geometry of the system.
 However, the fits were not robust owing to the unsolved degeneracies between a P.A. jump being either a phase wrap, or an orthogonal polarisation mode (OPM) transition or both (See also the conclusion chapter in  \citealt{VenkatramanKrishnanPhDThesis} for the temporally evolving P.A. profiles). Recently, \cite{VenkatramanKrishnanEtAl2019} used the total intensity and polarisation profile evolution to deduce that our line of sight has crossed the magnetic axis of the pulsar, and predicted that the pulsar will disappear from our line of sight before 2023.

Using the high resolution search mode acquisition capability of the MeerTime backend \citep{BailesEtAl2020}, we obtained 2048 seconds of search mode data, coherently dedispersed at the best DM of the pulsar, with full polartisation information and a high time resolution of $9\mu$s on April 8th 2020. The data were folded at the best known topocentric period of the pulsar with 8192 bins across its rotational phase, after which it was calibrated for polarisation and scrunched in frequency and time. Figure \ref{fig:J1141-6545} shows the pulsar polarisation profile and the P.A. swing. It can be seen from the figure that a 90 degree phase jump of the P.A. swing occurs, accompanied by a sudden dip in the linear polarisation. Combined, these show that the jump is indeed an OPM transition, solving the degeneracy faced by the earlier analysis with lesser S/N data from the Parkes telescope. This deduction can be fed back as a prior information to the full precessional RVM analysis to further understand its geodetic precession. As discussed in Section~\ref{sec:rvm}, such a flat P.A. curve is in principle difficult to describe within the RVM, but its changes with time (especially in the measured absolute value of the P.A.) due to relativistic spin precession  still provides valuable additional information.

\subsubsection{PSR J1756$-$2251}

\begin{figure}
    \centering
    \includegraphics[width=0.4\textwidth]{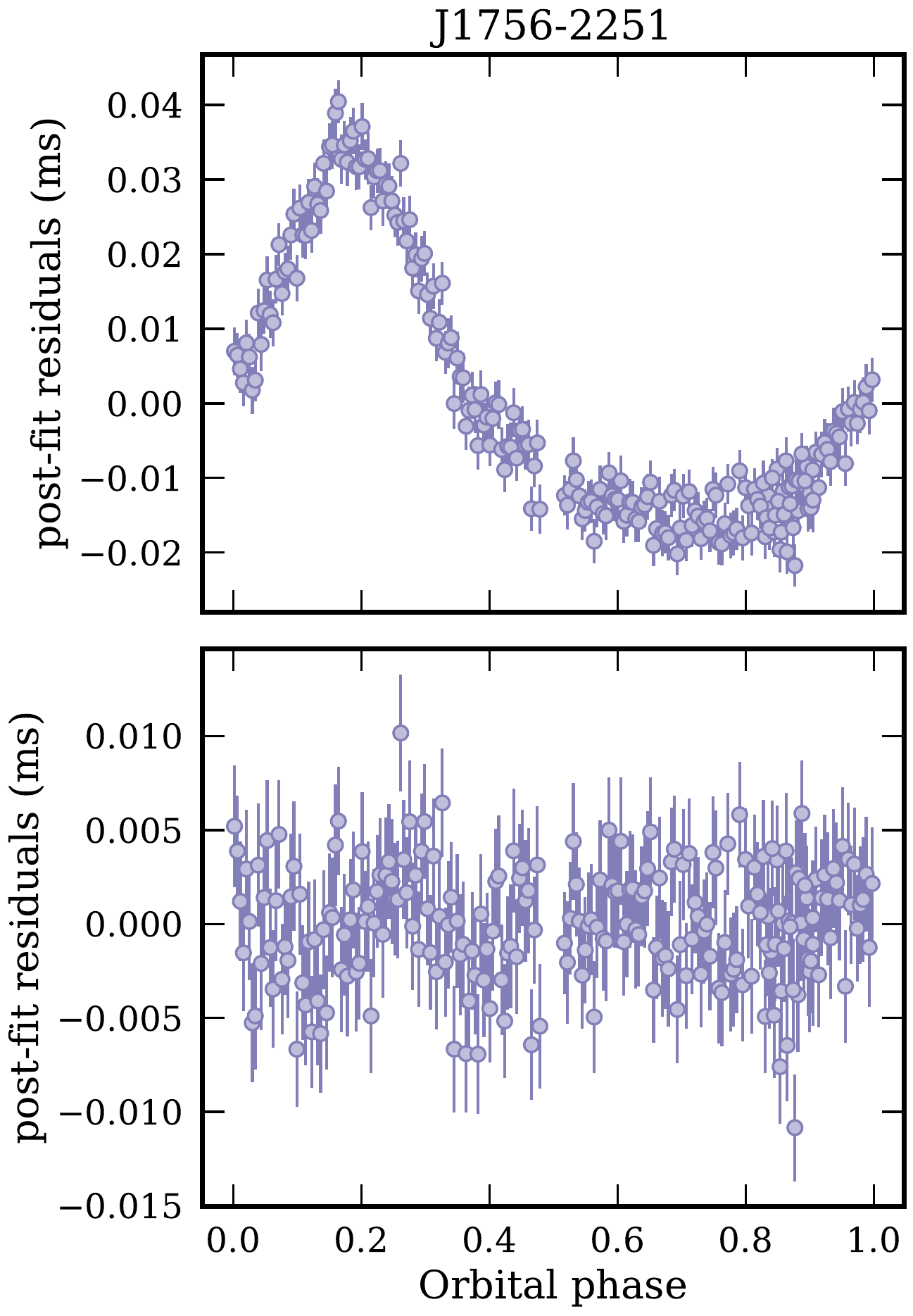}
    \caption{
    Shapiro delay signature in PSR J1756$-$2251 observed in a single full-orbit observation. The top panel shows the total un-absorbed Shapiro delay present in the data. The bottom panel shows the post-fit residuals after fitting for both the Keplerian orbital parameters and the Shapiro delay. The other parameters in all the panels are held fixed at their best value from \citet{FerdmanEtAl2014}, which explains the residual structure in the ToAs.  The RMS of the residuals in the top panel is $\sim16.9~\mu s$. Fitting for the Keplerian orbital parameters absorb part of the Shapiro delay leading to an improvement in the RMS to $\sim 4~\mu s$. Fitting for the Shapiro delay parameters further improves the RMS to $\sim 3.1~\mu s$ in the bottom panel. }
    \label{fig:J1756-2251}
\end{figure}

PSR J1756$-$2251 is a pulsar in a $7.7$-hour orbit around a companion star, most likely another neutron star. Since its discovery with the Parkes radio telescope \citep{FaulknerEtAl2004}, it has regularly been timed with a number of telescopes, resulting in the detection of several relativistic parameters. \cite{FerdmanEtAl2014} reported the latest timing results on this pulsar including the measurement of 5 relativistic parameters including the orbital period decay due to gravitational wave emission $\dot{P}_{\rm b}$ that was reportedly inconsistent with the value predicted by GR at the $2-\sigma$ confidence level. Motivated by this result, we had included this source in the RelBin programme.
The observations are already successful, as shown in Figure \ref{fig:J1756-2251} where we demonstrate  
the measurement of the Shapiro delay in the system with a single 8-hour long ($\sim$ full orbit) observation of the pulsar. This is in contrast to the Shapiro delay measurement presented by \cite{FerdmanEtAl2014} for their whole
available dataset (see their Figure 3). Holding other parameters fixed and fitting for spin period, DM, orbital phase, and the Shapiro delay PK parameters, we measure $r = 1.46 \pm 0.49 M_{\odot}$ and $s \equiv \sin i = 0.88 \pm 0.04$, of similar precision as and
consistent with \cite{FerdmanEtAl2014}.

\subsection{Sources for Mass Measurements}

As of November 2020, we have completed orbital campaigns for 10 pulsars, where we performed long observations over superior conjunction, followed by observations filling other parts of the orbit, and regular observations over the year to obtain a better timing baseline. Here, we only present some highlights for a subset of them, in order to demonstrate the ability of MeerKAT and the prospects of the RelBin programme.

\begin{figure}
    \centering
    \includegraphics[width=0.5\textwidth]{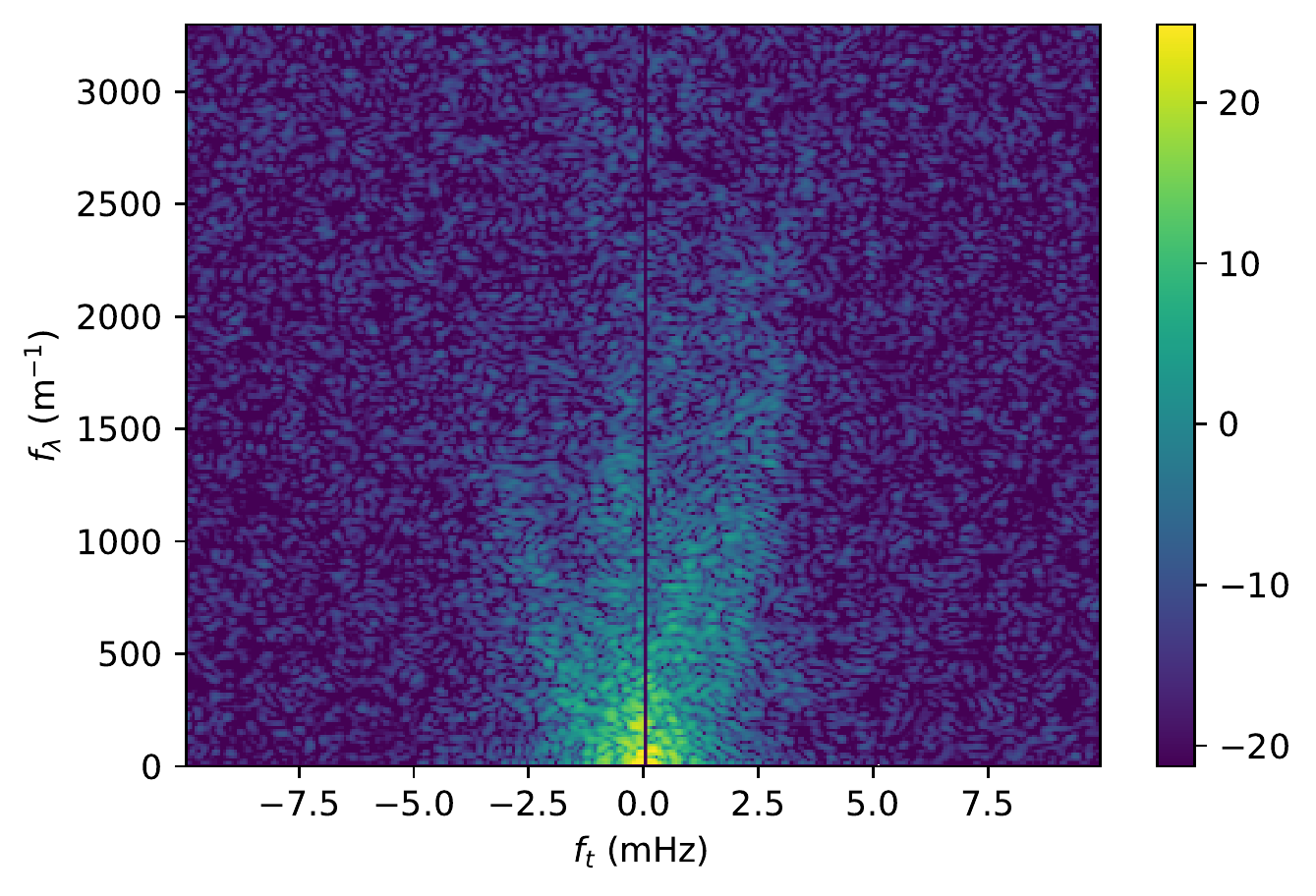}
    \caption{Secondary spectrum of J1757$-$5322 showing a clear scintillation arc, which we observe to change in curvature with the binary phase. The color scale represents the power (in arbitrary units) in each pixel. }
    \label{fig:1757_sspec}
\end{figure}

\subsubsection{PSR J1757$-$5322}
\label{sec:1757m53}
\label{sec:1757}

PSR J1757$-$5322 is a millisecond pulsar in a $11-$hour binary system around an optically identified white dwarf companion \citep{EdwardsAndBailes2001,JacobyEtAl2006}. The pulsar shows remarkable spectral features due to interstellar scintillation, owing to its proximity. This presents yet another avenue to measure some of the orbital parameters through scintillometry.

Here, we present 
secondary spectrum (Fourier transform of the dynamic spectrum) from a 90\,minute observation of this pulsar is shown in Figure \ref{fig:1757_sspec}. This reveals a parabolic scintillation arc. The degree of curvature for this arc depends on the distance to the source of scattering in the interstellar plasma, as well as the transverse velocity of the line of sight through this plasma. The dependence on velocity causes the arcs to change in curvature with the pulsar's binary velocity as a function of orbital phase. Measurements of this orbital modulation can be used to precisely determine $i$ and $\Omega$ (as in Reardon et al. \textit{submitted}). 
This measurement shows that we will 
be able to check the validity of our RVM results (see Section~\ref{sec:rvmfits}), even though initial timing efforts suggest that any Shapiro delay signal is weak. This result also demonstrates the good prospects of this method
for other RelBin sources, on which we will report later.

\begin{table}
    \centering
  \caption{Shown are the post-fitting model parameter values for PSR~J1811$-$2405 obtained with the DDFWHE timing model (\citealt{FreireAndWex2010}) using \textsc{Tempo}. The errors indicate nominal $\pm1~\sigma$ uncertainties.}
 \begin{minipage}{9cm}
\begin{tabular}{p{5.4cm}p{2.3cm}}
\hline
\hline
\multicolumn{2}{c}{Spin and astrometric parameters} \\
\hline
Right ascension, $\alpha$ (J2000) & 18:11:19.85405(3)\\
Declination, $\delta$ (J2000) & $-$24:05:18.41(2) \\
Proper motion in R.A., $\mu_{\alpha}$ (mas\,yr$^{-1}$) & 0.6(1) \\ 
Spin frequency, $\nu$ (Hz) & 375.856020042883244(7) \\
Spin down rate, $\dot{\nu}$~(s$^{-2}$)& -1.8895(3)$\times10^{-19}$ \\
Dispersion measure, DM (cm$^{-3}$\,pc) & 60.615(1) \\
Rotation measure (rad\,m$^{-2}$) & 30.3(2) \\
\hline
\multicolumn{2}{c}{Binary parameters}\\
\hline
Orbital period, $P_{\rm{orb}}$ (days) & 6.27230620515(7)\\
Projected semi-major axis, $x$ (lt-s) & 5.705656754(4) \\
Epoch of periastron, $T_{\rm{0}}$ (MJD) & 56328.98(2) \\
Longitude of periastron, $\omega$ ($^{\circ}$) & 62(1) \\
Orbital eccentricity, $e$ & 1.18(3)$\times10^{-6}$\\
Orthometric amplitude, $h_{3}$ ($\upmu$s) & 0.70(3) \\
Orthometric ratio, $\varsigma$ & 0.79(2) \\ 
\hline
\multicolumn{2}{c}{Derived parameters}\\
\hline
Companion mass from Bayesian analysis, $M_{\rm c}$ ($M_\odot$) & $0.29^{+0.04}_{-0.03}$ \\
Pulsar mass from Bayesian analysis, $M_{\rm p}$ ($M_\odot$)  & $1.8^{+0.4}_{-0.3}$ \\
Orbital inclination from Bayesian analysis, $i$  & $103{^\circ}.5^{+1{^\circ}.5}_{-1{^\circ}.9}$ \\
\hline
\multicolumn{2}{c}{Timing model} \\
\hline
Binary model & DDFWHE \\
Solar System ephemeris & DE435 \\
Reference epoch of period (MJD) & 56330.0 \\
Reference epoch of dispersion and rotation measure measurements (MJD)  & 58750.6 \\
First ToA (Rounded MJD) & 55871\\ 
Last ToA (Rounded MJD) & 58948 \\ 
Weighted RMS residuals ($\upmu$s) & 0.583 \\
Reduced $\chi^{2}$  & 0.9928\\
\hline \hline\label{tab:J1811-2405}
 \end{tabular}
\vspace{-0.5\skip\footins}
\end{minipage}
\end{table}

\subsubsection{PSR J1811$-$2405}

\label{sec:1811}

\begin{figure*}
    \centering
    \includegraphics[scale=0.55]{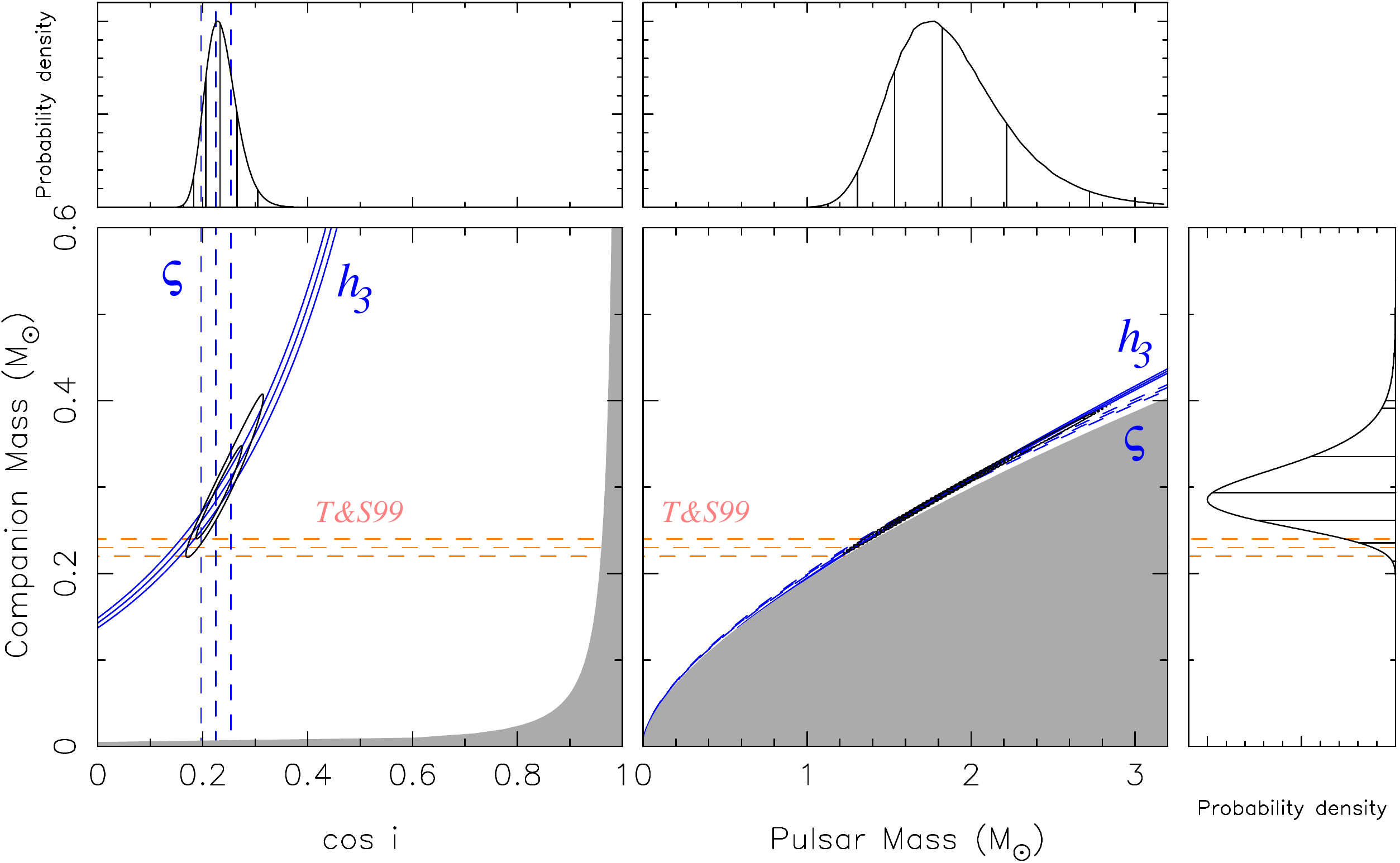}
    \caption{Mass and orbital inclination constraints for PSR J1811$-$2405. In the two main panels, the lines represent the nominal 68\% confidence intervals derived from the measured post-Keplerian parameters -- orthometric amplitude of the Shapiro delay ($h_3$) in solid blue and the orthometric ratio ($\varsigma$) in dotted blue \citep{FreireAndWex2010}. The solid black contours represent the 2-D probability distribution with their marginalized 1-D distributions represented in the top and right panels. The orange dotted lines indicate the nominal companion mass for the given orbital period as predicted by \citet{TaurisAndSavonije1999} for binary pulsars with Helium white dwarf companions.}
    \label{fig:mmp_J1811-2405}
\end{figure*}

PSR J1811$-$2405 was discovered by the HTRU survey \citep{KeithEtAl2010} conducted with the 64-m Parkes radio telescope. This is a pulsar in a $\sim 6.3$-day orbit around a likely Helium-WD companion. Since its discovery, it has been regularly timed with the Effelsberg and Nan\c{c}ay radio telescopes, providing a timing data span of almost 7 years. This resulted in the first detection of a Shapiro delay \citep{NgEtAl2020}. The detection, however, was marginal and resulted in large uncertainties on the pulsar mass, $M_{\rm p} = 2.0^{+0.8}_{-0.5} M_{\odot}$. We undertook observations of this pulsar with MeerKAT with the aim of obtaining a better detection of the Shapiro delay. We recorded a total of $\sim24$ hours, spanning $\sim14$ months, including a dense orbital campaign. During the orbital campaign, the pulsar was observed for 1 hour every day for a 7 day period ($\sim 1$ orbit) and was observed for 6 hours centred around superior conjunction. The data were reduced via standard techniques as mentioned in Section \ref{sec:obs_and_data_analysis} and the ToAs were analysed using \textsc{Tempo}
in unison with older data from \citet{NgEtAl2020}. 

Our results from the campaign are shown in Table \ref{tab:J1811-2405}, and the corresponding constraints on the component masses and orbital inclination can be seen in Fig. \ref{fig:mmp_J1811-2405}. 
Adding the much shorter timespan of our 
more precise RelBin data to most of those already presented by \cite{NgEtAl2020}, we obtain a 50\% improvement in the estimate the component masses compared to \cite{NgEtAl2020}, clearly demonstrating the contribution from the
MeerKAT observations.
Although the measurement of the MSP mass is not yet precise enough to be astrophysically interesting - the 1-$\sigma$ error bars includes all regions between $\sim\, 1.5$ and 2.1 $M_{\odot}$, we can obtain insight for 
the WD companion: 
its mass is only 2-$\sigma$ compatible with the prediction of \cite{TaurisAndSavonije1999} of $\sim \, 0.22 \, M_{\odot}$,
which assumes the companion is a He WD. If the WD companion really is much more massive than that, then
it might be a CO WD instead. In this case the MSP could also be substantially more massive than the $\sim \, 1.3 \, M_{\odot}$ that it would have if the companion mass were 0.22 $M_{\odot}$.

Future orbital campaigns of this system with MeerKAT with the UHF (where the pulsar is brighter) and the S-band (where the pulse profile is known to have narrow features) receivers have the potential to improve the current estimates of the masses, and
resolve both the issue of the nature of the WD companion and provide a more precise estimate of the pulsar mass.

We use this pulsar also to demonstrate how an RVM fit discussed in Section~\ref{sec:rvm} can be used to break the $\sin i$ ambiguity of the Shapiro delay measurement. We point out that \cite{NgEtAl2020} already obtained an RVM fit to Parkes data. Our data and solution presented here are more precise but fully consistent with the earlier result. As shown in Table~\ref{tab:rvm}, our RVM fit done with a uniform prior for $\zeta$ over the full 0 to 180 deg range, results in an estimate for an orbital inclination angle of $i=104.0(3)$ deg. This does not only break the $\sin i$ ambiguity but is also in perfect agreement with the Shapiro delay measurement in numerical value. This is indeed both the case when fitting for a non-zero $\Delta$ parameter as in Eqn.~\ref{eqn:rvm} (obtaining $\Delta = 7.5\pm0.5$ deg), or without it (i.e.~setting $
\Delta=0$, with entries as in Tab.~\ref{tab:rvm}). Consequently, we list the inclination angle in Table~\ref{tab:J1811-2405}
as $i=103{^\circ}.5^{+1{^\circ}.5}_{-1{^\circ}.9}$ deg.

\begin{figure*}
    \centering
    \includegraphics[scale=0.45]{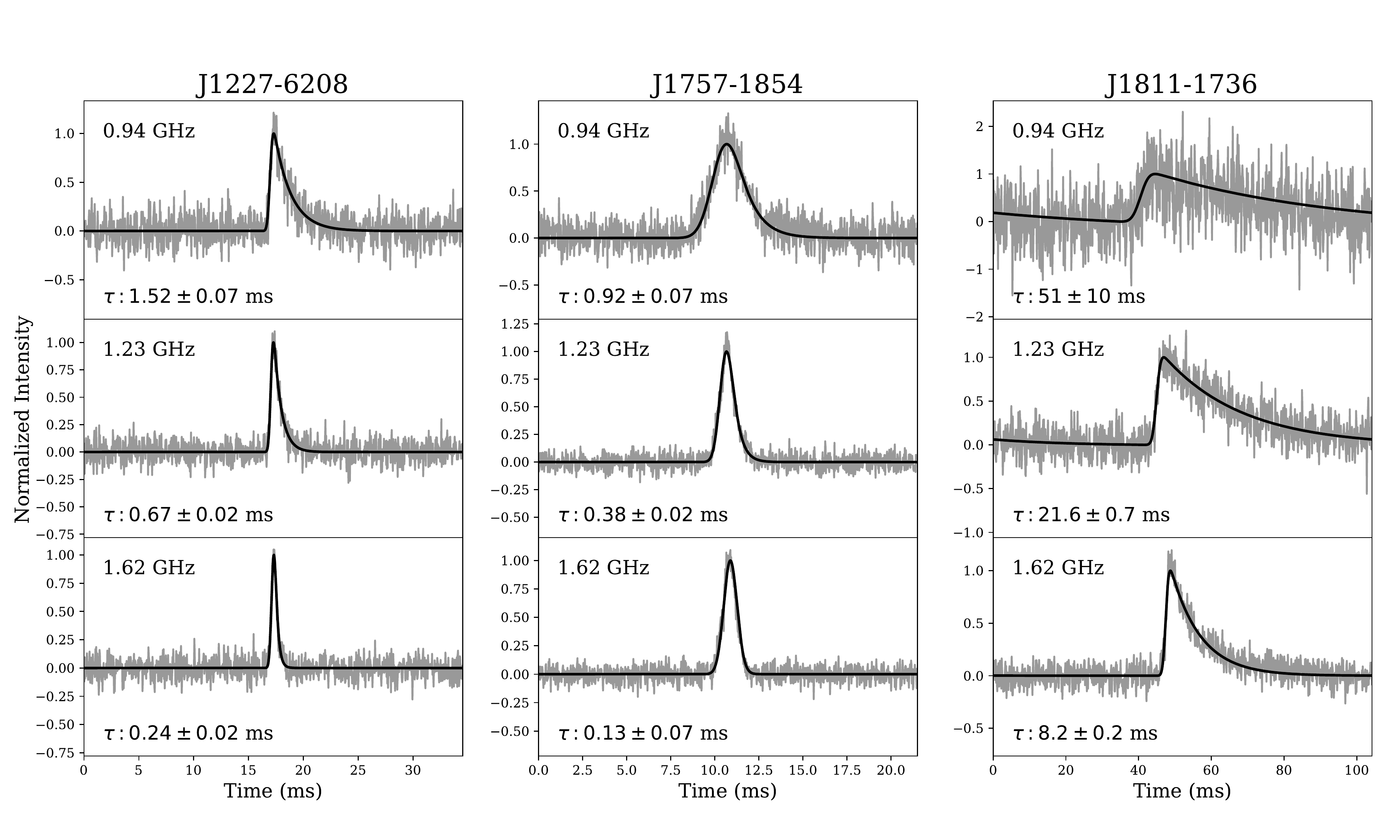}
    \caption{Measurements of scattering timescales at three selected sub-bands for PSRs J1227$-$6208 (left), J1757$-$1854 (middle) and J1811$-$1736 (right). The gray background lines denote the actual data, and the solid black lines denote the fits to the data with a scattering model. }
    \label{fig:scatter_profiles}
\end{figure*}

\begin{figure}
    \centering
    \includegraphics[scale = 1.3]{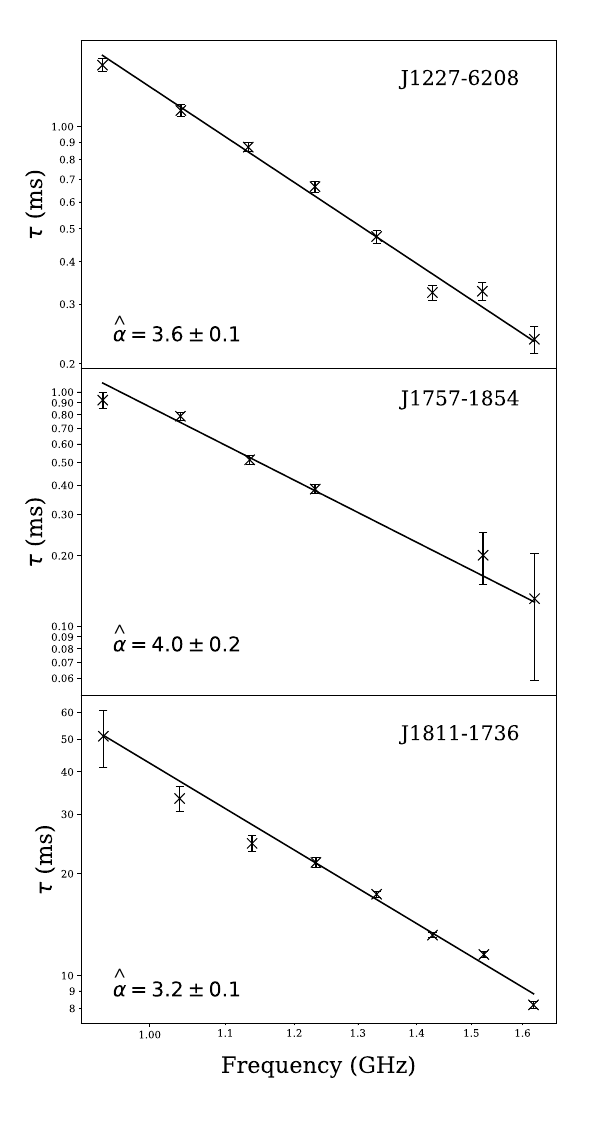}
    \caption{Measured scattering times measured for PSRs J1227-6208, J1757$-$1854 and J1811-1736 across the observing bandwidth at L-band and the derived corresponding spectral index.}
    \label{fig:SI_fits}
\end{figure}

\section{Further observing strategy \& Outlook}

\label{sec:outlook}

Our initial results presented above suggest that MeerKAT will provide superior data compared to those already available from less sensitive telescopes. This includes wide-bandwidth full-polarisation data of the observed pulse profiles as well as high-precision timing data. We have used ``pulse structure'' data  to obtain additional information that is useful for undertaking tests of relativistic gravity, as was first suggested by \cite{DamourTaylor1992}. We have also shown that the simultaneous dynamic spectra obtained over large bandwidths yield yet another dimension that can be explored for our science goals. Finally, the well-calibrated, high S/N pulse profiles allow us to obtain a measured timing precision, which scales with both increased S/N as well as our ability to resolve narrow pulse features \citep[e.g.][]{Lorimer&Kramer2005}. Despite these impressive improvements available with MeerKAT we point out that the already existing data sets obtained with other telescopes will remain an essential input in the timing analysis presented in dedicated publications elsewhere, as they provide the long timing baseline that is essential to measure secular PK parameters. 

We have demonstrated the performance of MeerKAT at L-Band and UHF frequencies; the RelBin programme will also make excellent use of the upcoming S-band system.  Some pulsars have been found to have improved timing precision at these higher frequencies (e.g.~PSR J1757$-$1854, \citealt{CameronEtAl2018}), where the effects of the interstellar medium are markedly reduced. This includes the impact of variations of the dispersion measure, but in particular the reduced interstellar scattering effects on some pulsars. Here, we draw attention to three sources on our initial priority’s  list, i.e. PSRs J1227$-$6208 and J1811$-$1736, the profiles of which are significantly scattered at L-band (see Figures~\ref{fig:flatPAa} and \ref{fig:complexPAa}, respectively), as well as the mentioned PSR J1757$-$1854 (Fig.~\ref{fig:flatPAb}).  In order to gauge the impact of high-frequency observations, we have
performed a scattering analysis for these pulsars by splitting the data into 8 sub-bands, which we describe in the following.

In each sub-band, we fit for a time series scattering model, as described in \citet{GeyerEtAl2017}. We used the affine invariant MCMC sampling algorithm implemented in the {\sc emcee} software \citep{EMCEE} to sample the log-likelihood function over the model parameter space, imposing flat priors that constrain the parameter space only to physically possible values. Full details of the modelling process are found in Oswald et al.~(to be submitted). For pulse width measurements, we used analytical pulse profiles that are obtained from a Gaussian process (GP) non-parametric model of the Stokes I data in each frequency band. Based on the GP-determined noise level (rms), we required a signal-to-noise ratio of at least 3 at the location of the respective width level (e.g., at 50\% of the pulse maximum for W50) for a measurement to be made. 
Further details of the pulse width measurement technique can be found in Posselt et al.~(in prep.).  The results for the three pulsars at selected sub-bands can be seen in Fig.~\ref{fig:scatter_profiles}. The decrease in width across the frequencies is clearly visible, most strikingly for PSR J1811$-$1736, but also PSR J1757$-$1854 becomes clearly narrower which explains the better timing precision. 

One can fit a power law to the frequency dependence of the measured scattering times, i.e.~$\tau \propto f^{-\hat{\alpha}}$. The results of those fits is shown in Fig.~\ref{fig:SI_fits}. The spectral indices, $\hat{\alpha}$, 
found are all near $-4$ as expected from the turbulent ISM \citep{Lorimer&Kramer2005}.
We can use these spectral-index fits to predict the timing precision for the three pulsars at S-band. Doing so, we can expect the timing precision to  reach $\lesssim1 \mu$s precision for observations with the upcoming S-band system. Even for PSR~J1811$-$1736, the expected scattering timescale for frequencies above 3 GHz will be less than 1 $\mu$s. This result is consistent with  \cite{CorongiuEtAl2007}, who also measured the same scattering index of $-3.5(1)$ and already suggested that timing at higher frequencies will provide superior precision of the timing parameters. With this in mind, we expect to be able to make an improved measurement of $\dot\omega$ and a first measurement of the Shapiro delay, and potentially the PK parameter $\gamma$ for PSR J1811$-$1736. In the case of PSR J1227$-$6208 we expect a detection of a Shapiro delay that will give us access to the system's masses. Our fits also confirm that PSR J1757$-$1854 will perform much better at S-Band.

As data for RelBin are collected, the timing precision will continue to improve with time. The rate of improvement for PK parameters varies depending on the parameter \citep{DamourTaylor1992,Lorimer&Kramer2005}, while existing data will provide an important head start in all our cases. How a combination of existing and MeerKAT data will lead to further discoveries and insight is demonstrated for PSR J0737$-$3039A/B by \cite{huhu2020}. We refer to the latter work as a case study. 

\section{Summary \& Conclusions}

We presented the science case and initial results from the MeerTime relativistic binary programme (RelBin) with the MeerKAT telescope. Our observations demonstrate that MeerKAT is an powerful pulsar instrument, and is capable of delivering high fidelity data that will yield tests of theories of gravity as well as a large sample of new or significantly improved mass measurements. This is made possible by superior timing precision in combination with additional information derived from polarisation profiles and dynamic spectra. We demonstrated this by presenting a rare collection of well calibrated polarisation profiles of suitable pulsars and their analysis. We demonstrate the utility of dynamic spectra to determine system geometries by modelling changes in the scintillation properties. Full-polarisation information is also crucial for achieving the anticipated timing precision \citep{vanStraten2006}. Hence, it is important to also confirm that the polarisation properties of the instrument are consistent across the observing bands, while the long term timing stability ensures high precision pulsar timing. Our observations through the lifetime of MeerTime hence have the potential to significantly contribute to our knowledge of neutron star masses and further understanding gravity. Finally, we indicated the potential of high-precision timing observations with the S-band system. We will report on first results with this system when available. With the initial results presented here, we are confident that RelBin can achieve the science goals it has set out to attain.

\section*{Acknowledgements}
The MeerKAT telescope is operated by the South African Radio Astronomy Observatory, which is a facility of the National Research Foundation, an agency of the Department of Science and Innovation. SARAO acknowledges the ongoing advice and calibration of GPS systems by the National Metrology Institute of South Africa (NMISA) and the time space reference systems department department of the Paris Observatory. MeerTime data is housed on the OzSTAR supercomputer at Swinburne University of Technology. The Parkes radio telescope is funded by the Commonwealth of Australia for operation as a National Facility managed by CSIRO. This research has made extensive use of NASAs Astrophysics Data System (https://ui.adsabs.harvard.edu/) and includes archived data obtained through the CSIRO Data Access Portal (http://data.csiro.au). Parts of this research were conducted by the Australian Research Council Centre of Excellence for Gravitational Wave Discovery (OzGrav), through project number CE170100004 and the Laureate fellowship number FL150100148. The MeerTime Pulsar Timing Array acknowledges support of the Gravitational Wave Data Centre funded by the Department of Education via Astronomy Australia Ltd. and ADACS.  MBu, APo, and AR used resources from the research grant “iPeska” (P.I. Andrea Possenti) for this work, funded under the INAF national call Prin-SKA/CTA approved with the Presidential Decree 70/2016.  RMS acknowledges support through Australian Research Council fellowship FT190100155. LO acknowledges funding from the UK Science and Technology Facilities Council (STFC) Grant Code ST/R505006/1. MK, VVK, PCCF, FA, DJC, TG, AP, NW, and EDB acknowledge continuing valuable support from the Max-Planck Society.


\section*{Data Availability}
The data underlying this article will be shared on reasonable request to the corresponding author.



\bibliographystyle{mnras}
\bibliography{relbin,global} 


\bsp	
\label{lastpage}
\end{document}